\documentclass[iop,revtex4,letterpaper,twocolappendix]{emulateapj}
\usepackage{natbib}
\usepackage[colorlinks=true,urlcolor=blue,citecolor=blue,linkcolor=blue,breaklinks]{hyperref}
\usepackage{amsmath}
\usepackage{pdflscape}
\usepackage{rotate}

\begin{document}

\newcommand{\coto}{\mbox{CO(2--1)}}
\newcommand{\od}{$\tau_{\rm{353\,GHz}}$}
\newcommand{\pao}{$\rm{PA_{Out}}$}
\newcommand{\paf}{$\rm{PA_{Fil}}$}
\newcommand{\pafilf}{$\rm{PA_{Fil,F}}$}
\newcommand{\gaf}{$\gamma_{\rm{_{F}}}$}
\newcommand{\pafone}{$\gamma_{1\arcmin}$}
\newcommand{\paftwo}{$\gamma_{2\arcmin}$}
\newcommand{\pafthree}{$\gamma_{3\arcmin}$}
\newcommand{\paffour}{$\gamma_{4\arcmin}$}
\newcommand{\paffive}{$\gamma_{5\arcmin}$}
\newcommand{\pafsix}{$\gamma_{6\arcmin}$}
\defcitealias{Anathpindika2008}{AW08}	

\title{Alignment Between Protostellar Outflows and Filamentary Structure} 

\author{Ian W. Stephens\altaffilmark{1},  Michael M. Dunham\altaffilmark{2,1}, Philip C. Myers\altaffilmark{1}, Riwaj Pokhrel\altaffilmark{1,3}, Sarah I. Sadavoy\altaffilmark{1}, Eduard I. Vorobyov\altaffilmark{4,5,6}, John J. Tobin\altaffilmark{7,8}, Jaime E. Pineda\altaffilmark{9}, Stella S. R. Offner\altaffilmark{3}, Katherine I. Lee\altaffilmark{1}, Lars E. Kristensen\altaffilmark{10}, Jes K. J\o rgensen\altaffilmark{11}, Alyssa A. Goodman\altaffilmark{1}, Tyler L. Bourke\altaffilmark{12}, H\'{e}ctor G. Arce\altaffilmark{13}, Adele L. Plunkett\altaffilmark{14}}

\altaffiltext{1}{ Harvard-Smithsonian Center for Astrophysics, 60 Garden Street, Cambridge, MA, USA
ian.stephens@cfa.harvard.edu}
\altaffiltext{2}{Department of Physics, State University of New York at Fredonia, 280 Central Ave, Fredonia, NY 14063, USA}
\altaffiltext{3}{Department of Astronomy, University of Massachusetts, Amherst, MA 01003, USA}
\altaffiltext{4}{Institute of Fluid Mechanics and Heat Transfer, TU Wien, Vienna, 1060, Austria}
\altaffiltext{5}{Research Institute of Physics, Southern Federal University, Stachki Ave. 194, Rostov-on-Don, 344090, Russia}
\altaffiltext{6}{University of Vienna, Department of Astrophysics, Vienna, 1180, Austria}
\altaffiltext{7}{Homer L. Dodge Department of Physics and Astronomy, University of Oklahoma, 440 W. Brooks Street, Norman, OK 73019, USA}
\altaffiltext{8}{Leiden Observatory, Leiden University, P.O. Box 9513, 2300-RA Leiden, The Netherlands}
\altaffiltext{9}{Max-Planck-Institut f\"ur extraterrestrische Physik, D-85748 Garching, Germany}
\altaffiltext{10}{Centre for Star and Planet Formation, Niels Bohr Institute and Natural History Museum of Denmark, University of Copenhagen, \O ster Voldgade 5-7, DK-1350 Copenhagen K, Denmark}
\altaffiltext{11}{Niels Bohr Institute and Center for Star and Planet Formation, Copenhagen University, DK-1350 Copenhagen K., Denmark}
\altaffiltext{12}{SKA Organization, Jodrell Bank Observatory, Lower Withington, Macclesfield, Cheshire SK11 9DL, UK}
\altaffiltext{13}{Department of Astronomy, Yale University, New Haven, CT 06520, USA}
\altaffiltext{14}{European Southern Observatory, Av. Alonso de Cordova 3107, Vitacura, Santiago de Chile, Chile}
\interfootnotelinepenalty=10000

\begin{abstract}
We present new Submillimeter Array (SMA) observations of CO(2--1) outflows toward young, embedded protostars in the Perseus molecular cloud as part of the Mass Assembly of Stellar Systems and their Evolution with the SMA (MASSES) survey.  For 57 Perseus protostars, we characterize the orientation of the outflow angles and compare them with the orientation of the local filaments as derived from $Herschel$ observations. We find that the relative angles between outflows and filaments are inconsistent with purely parallel or purely perpendicular distributions.  Instead, the observed distribution of outflow-filament angles are more consistent with either randomly aligned angles or a mix of projected parallel and perpendicular angles. A mix of parallel and perpendicular angles requires perpendicular alignment to be more common by a factor of $\sim$3. Our results show that the observed distributions probably hold regardless of the protostar's multiplicity, age, or the host core's opacity. These observations indicate that the angular momentum axis of a protostar may be independent of the large-scale structure. We discuss the significance of independent protostellar rotation axes in the general picture of filament-based star formation.
\end{abstract}

\subjectheadings{stars: formation -- galaxies: star formation -- stars: protostars -- ISM: jets and outflows  -- ISM: clouds -- ISM: structure}

\maketitle

\section{Introduction} Ê\label{sec:intro}
Many stars form in filamentary structures with widths of order 0.1 pc \citep[e.g.,][]{Arzoumanian2011}. While the exact shape of filaments is debated, e.g., cylinders versus ribbons \citep{Auddy2016}, filaments are defined by a long axis and two much shorter axes. Dense cores ($\sim$0.1\,pc scale) either form within the filaments or form simultaneously with the filaments \citep{ChenOstriker2015}. Inhomogeneous flow or shear from colliding flows can torque cores \citep[e.g.,][]{Fogerty2017,Clarke2017}. Classically, angular momentum is expected to be hierarchically transferred from molecular clouds to cores to protostars \citep[e.g.,][]{Bodenheimer1995}. For a star-forming filament, large-scale flows are probably either onto the short axes of the filament from its cloud (either via accretion from the cloud or accretion via a collision) or along the long filamentary axis. In a simplistic, non-turbulent scenario where one of the flows about the three filamentary axes dominates, a core will likely rotate primarily parallel or perpendicular to the parent filament. If the angular momentum direction at the protostellar scale is inherited from this core scale, the rotation axes of newly formed protostars will also be preferentially parallel or perpendicular to the filaments.



 One way to empirically test the alignment between a protostar's spin and its filamentary structure is to observe a protostar's outflow direction and compare it to the filamentary structure as probed by dust emission. By using this method across five nearby star-forming regions, \citet{Anathpindika2008} found suggestive evidence that outflows (as traced by scattered light) tend to be preferentially perpendicular to filaments. On the other hand, \citet{Davis2009} found that in Orion, the orientation between outflows (as traced by H$_2$) and filaments appear random. A well-focused study that analyzes the outflow-filament angles is needed to reconcile this disagreement.

The rotation axis of a protostar, or even the parent protostellar core, could also be independent of its natal filamentary structure. Some observations have shown that the angular momentum vectors of cores themselves may be randomly distributed about the sky, regardless of the cloud, core, or filamentary structure \citep{Heyer1988,Myers1991,Goodman1993,Tatematsu2016}. Multiplicity could also affect rotation axes. In the Submillimeter Array \citep[SMA,][]{Ho2004} large project called the Mass Assembly of Stellar Systems and their Evolution with the SMA (MASSES; co-PIs: Michael Dunham and Ian Stephens), \citet{Lee2016} found that outflows of wide-binary pairs (i.e., binary pairs separated by 1000\,AU and 10,000\,AU) are typically randomly aligned or perpendicular (but not parallel) to each other. Radiation-magnetohydrodynamic simulations by \citet{Offner2016} of slightly magnetically-supercritical turbulent cores found the same results for wide-binary pairs. These simulations suggest that the direction of the protostellar spin axis can evolve significantly during formation, indicating that, at least for wide-binaries, the rotation axes are independent of the large-scale structure. 




In this paper, we aim to observationally test whether or not a preferential alignment exists between the local filamentary elongation and the angular momentum axis as traced by outflows. To test such alignment, we use new CO observations from the MASSES survey to trace the molecular outflows in the Perseus molecular cloud. Along with ancillary data, we determine accurate projected outflow position angles (PAs) for 57 Class 0 and I protostars. The MASSES survey provides uniform spatial coverage of the same molecular line tracers in a single cloud, and only focuses on young sources -- Class 0 and I protostars. Since these protostars are young, their parent filamentary structure has had less time to change in morphology since the birth of the stars. These outflow observations can then be compared to the filamentary structure as observed by the $Herschel$ Gould Belt survey \citep[e.g.,][]{Andre2010}.
 
We describe the observations used in Section~\ref{observations} and the outflow/filament PA extraction techniques in Section~\ref{dataanalysis}. We present the results in Section~\ref{results} and discuss their possible implications in Section~\ref{discussion}. Finally, we summarize the main results in Section~\ref{summary}.


\section{Observations }\label{observations}

\subsection{Outflow and Continuum Data}
For the Perseus protostellar outflows studied in this paper, we introduce new, unpublished MASSES CO(2--1) data. The SMA observations were calibrated using the MIR software package\footnote{\url{https://www.cfa.harvard.edu/~cqi/mircook.html}} and imaged using the MIRIAD software package \citep{Sault1995}.  More details of the data reduction process for the MASSES survey are presented in \citet{Lee2015}. The new MASSES data all come from the SMA's subcompact configuration, which typically has baselines between 3\,k$\lambda$ and 54\,k$\lambda$, resulting in an average synthesized beam size of $\sim$3$\farcs$8. The velocity resolution of the observations is 0.26\,km\,s$^{-1}$, and the data were smoothed to 0.5\,km\,s$^{-1}$ in this study. The typical 1$\sigma$ rms in a 0.5\,km\,s$^{-1}$ channel is 0.15\,K.

Along with the new MASSES CO(2--1) data, we also used new MASSES 1.3\,mm continuum data to locate the centroid of the bipolar outflow, which is used to help measure the outflow PAs (see Section~\ref{outflowPA}). A more detailed analysis of the continuum data will be discussed in a forthcoming paper (R.~Pokhrel et al. in preparation). The SMA data will become publicly available with the MASSES data release paper (I.~Stephens et al. in preparation).

In some cases, we use already published CO PAs (primarily from \citealt{Plunkett2013} and from other MASSES data published in \citealt{Lee2015,Lee2016}) since these observations were either better quality and/or at higher resolution. These published PAs each came from observations of one of three different $J$ rotational transitions of CO: CO(1--0), CO(2--1), and CO(3--2). The rest frequencies for these three spectral lines are 115.27120\,GHz, 230.53796\,GHz, and  345.79599\,GHz, respectively.


\subsection{$Herschel$-derived Optical Depth Maps}\label{opticaldepth}
$Herschel$ is well-suited for finding filaments in Perseus given its resolution and wavelength range. The resolution at the longest $Herschel$ wavelength (500\,$\mu$m) is 36$\arcsec$ or $\sim$0.04\,pc at the distance of Perseus \citep[235\,pc,][]{Hirota2008}. Star-forming filaments have temperatures of $\sim$10 to 20\,K, and thus the dust continuum will peak within the $Herschel$ bands (70\,$\mu$m to 500\,$\mu$m). These wavebands can be used to approximate the optical depth and the column density of Perseus filaments. Indeed, several studies have already created optical depth or column density maps of the Perseus molecular cloud using $Herschel$ observations, including \citet{Sadavoy2014}, \citet{Zari2016}, and \citet{Abreu-Vicente2016}. All three of the aforementioned studies assumed a modified blackbody with a specific intensity of 
\begin{equation}
I_\nu = B_\nu(T)(1-e^{-\tau_\nu}) \approx B_\nu(T)\tau_\nu ,
\end{equation}
where $B_\nu$ is the blackbody function at temperature $T$ and $\tau_\nu$ is the optical depth. $\tau_\nu$ is assumed to follow a power-law function of the form $\tau_\nu \propto \nu^\beta$, where $\beta$ is the dust emissivity index. The dust column density, $N_{\rm{dust}}$, can be calculated assuming $\tau_\nu = N_{\rm{dust}} \kappa_\nu$, where $\kappa_\nu$ is the dust opacity. Each study assumed $\tau_\nu$ and $T$ to be free parameters.

\begin{figure*}[ht!]
\begin{center}
\includegraphics[width=1.9\columnwidth]{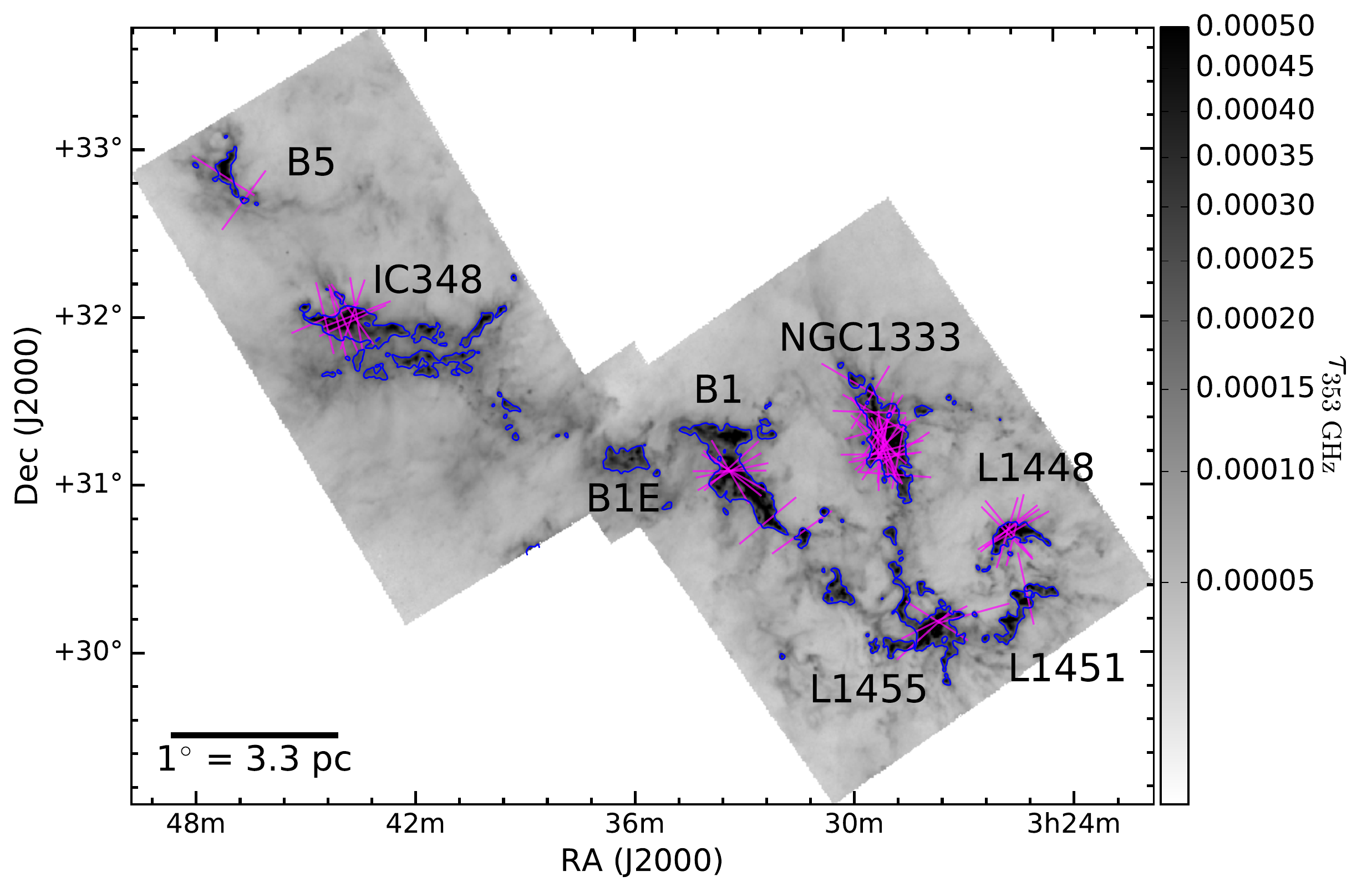}
\end{center}
\caption{\od\ map of the Perseus molecular cloud \citep{Zari2016}, with magenta lines showing the directions of the outflows measured in this study. The size of the lines only represents the direction of the outflow and not the angular extent. Thin blue contours are shown for \od~=~0.0002. These contours roughly show the boundaries of each labeled clump and correspond to a column density of $N$(H$_2$)~$\approx$~5~$\times$10$^{21}$\,cm$^{-2}$ \citep{Sadavoy2014}.
}
\label{all_tau_outflows} 
\end{figure*}

While these studies varied slightly, e.g., on their assumption for $\beta$, the resulting maps are very similar. We choose to use the 353~GHz optical depth (\od) map from \citet{Zari2016} since this map has been made publicly available. \citet{Zari2016} assumed a value of $\beta$~=~2, and they did not convert the \od\ maps to column density. The \od\ maps were made using only the $Herschel$ 160, 250, 350, and 500\,$\mu$m maps. Each $Herschel$ map was zero-point corrected with $Planck$ and smoothed to the coarsest resolution (500\,$\mu$m), resulting in an \od\ map at 36$\arcsec$ resolution.  The final \od\ map has the pixels regridded to equatorial coordinates with pixel sizes of $18\arcsec\times18\arcsec$. This \od\ map also includes coarse resolution $Planck$ \od\ maps in the field external to the $Herschel$ observations. 

Figure~\ref{all_tau_outflows} shows the \citet{Zari2016} \od\ map of Perseus.  For simplicity, we masked out the $Planck$-only regions of the map which extend beyond the $Herschel$ observations.  The resolution of these $Planck$-only regions are too coarse to resolve the filaments and none of our MASSES targets are located within them.


\begin{figure*}[ht!]
\begin{center}
\includegraphics[width=2\columnwidth]{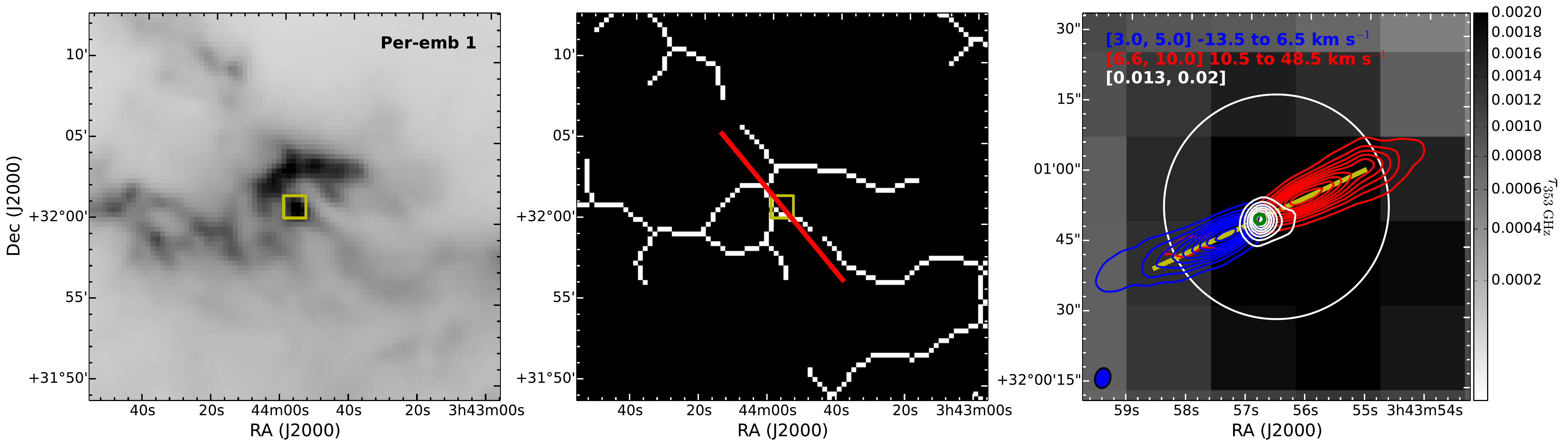}
\includegraphics[width=2\columnwidth]{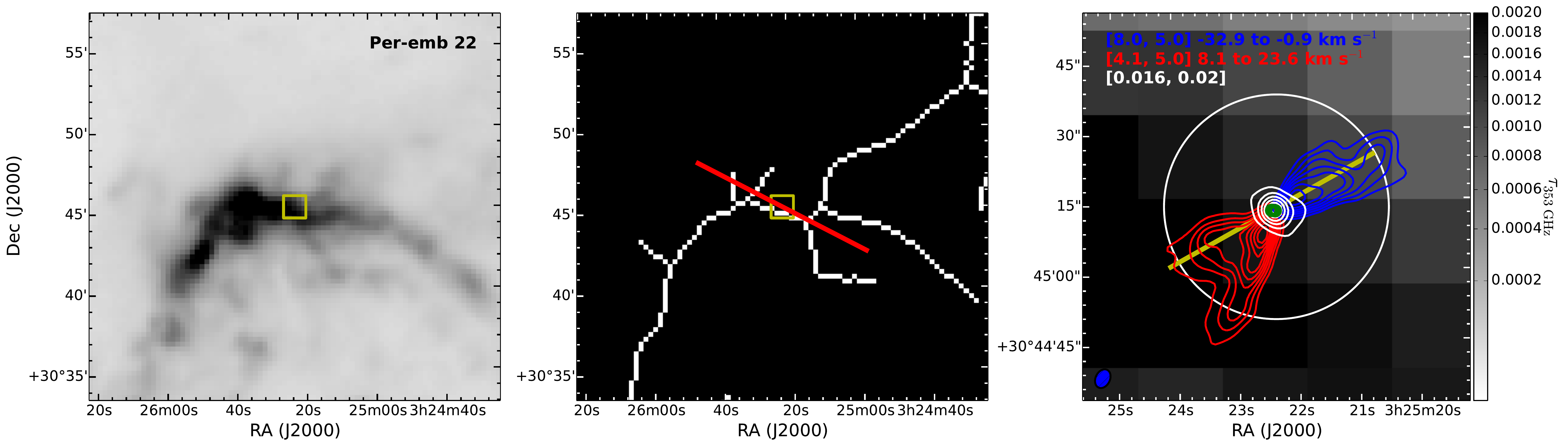}
\includegraphics[width=2\columnwidth]{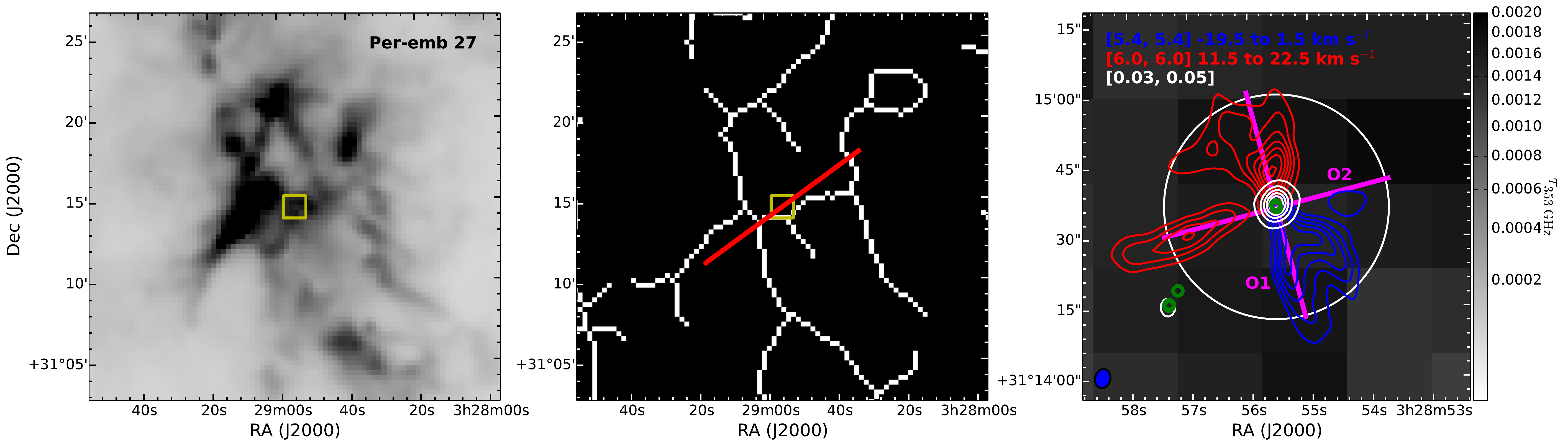}
\end{center}
\caption{Figures demonstrating the \texttt{FILFINDER} algorithm for Per-emb~1 (top 3 panels), Per-emb~22 (middle 3 panels), and Per-emb~27 (bottom 3 panels); other Perseus protostars can be found in the eletronic version of this paper. The left and middle panels show the \od\ maps \citep{Zari2016} and the fitted filament skeletons from \texttt{FILFINDER} \citep{Koch2015}, respectively. The red line in the middle panel shows the fitted \pafilf\ for the protostar. The yellow squares in these two panels show the area we zoom-in for the right panels. The right panels show the \od\ overlaid with SMA red and blue CO(2--1) integrated intensity contours of the red and blue lobes, respectively. The white contours show the SMA 1.3\,mm continuum. The color-coded bracketed numbers in the top left give the first contour level followed by the contour level increment for each subsequent contour. The CO(2--1) contour levels and increments are in units of Jy\,beam$^{-1}$\,km\,s$^{-1}$ while the continuum contour levels and increments are in units of Jy\,beam$^{-1}$. The red and blue velocity interval for CO(2--1) intensity integration are shown next to their corresponding contour levels.  The small green circles show the location of the protostellar sources as determined at high resolution by the VLA \citep{Tobin2016}. The measured \pao\ is shown as a line under the contours, and the line is yellow if \pao\ comes from this study, and magenta if \pao\ comes from other studies (as indicated in Table \ref{tab:angles}). The white circle shows the 48$\arcsec$ diameter (FWHM) primary beam of the SMA.
}
\label{filament_fit} 
\end{figure*}

\begin{figure*}[ht!]
\begin{center}
\includegraphics[width=2\columnwidth]{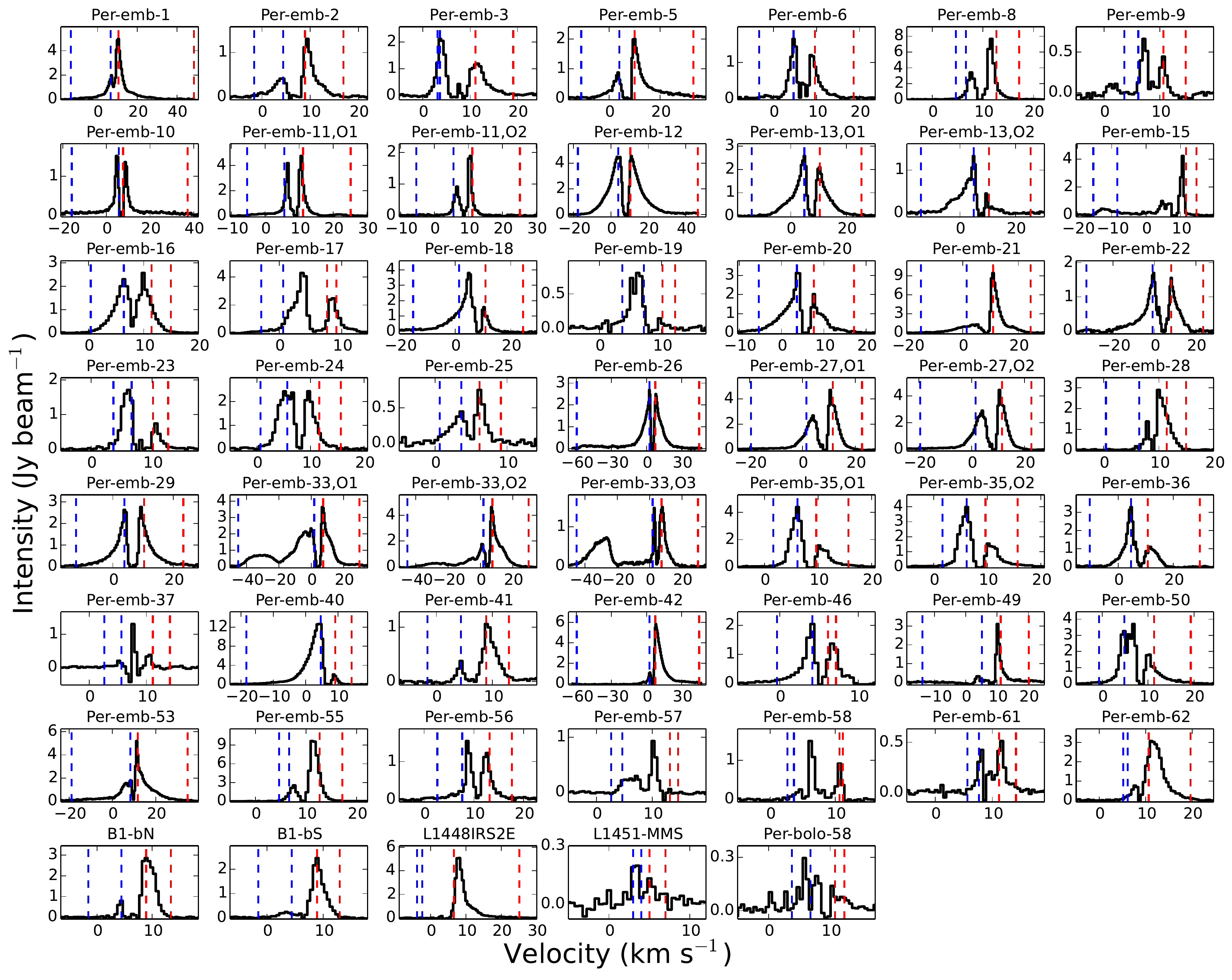}
\end{center}
\caption{Average CO(2--1) spectra within a radius of 8$\arcsec$ from each protostar, where the protostar's position is given in Table~\ref{tab:angles}. The velocity resolution is 0.5\,km\,s$^{-1}$. The vertical dashed lines show the interval ranges used to produce the integrated intensity maps in the right panels of Figure~\ref{filament_fit}. The two blue and two red lines show the integrated intervals for the blue- and red-shifted emission, respectively. These integrated intensity ranges were manually adjusted to produce the best visualization of the outflows for each source. In some cases, no outflows were found for a particular lobe, or the lobe emission was difficult to extract from the large-scale CO(2--1) emission. Note that for Per-emb-57, the dominant outflow emission is toward the southeast, more than 8$\arcsec$ from the source's center, and thus the spectrum poorly represents the outflow emission.
}
\label{spectra} 
\end{figure*}

\section{Data Analysis Techniques} \label{dataanalysis}
In this section, we summarize how we measure PAs for both outflows and filaments from observations. All angles are measured counterclockwise from the north celestial pole. These PAs are used to calculate the main parameter of interest, $\gamma$, which is the projected angle difference between the outflows and filaments. Specifically, $\gamma$ is given by  
\begin{equation}
\gamma = \rm{MIN}\big\{\rm{|PA_{Out}}-PA_{Fil}|,180^\circ-\rm{|PA_{Out}}-PA_{Fil}|\big\},
\end{equation}\label{filequation}
where $\rm{PA_{Out}}$ and $\rm{PA_{Fil}}$ are the PAs of the outflow and filament, respectively. MIN indicates that we are interested in the minimum of the two values in the brackets. Table~\ref{tab:angles} lists the measured PAs for all outflows and filaments in this study.  


\begin{figure*}[ht!]
\begin{center}
\includegraphics[width=2.1\columnwidth]{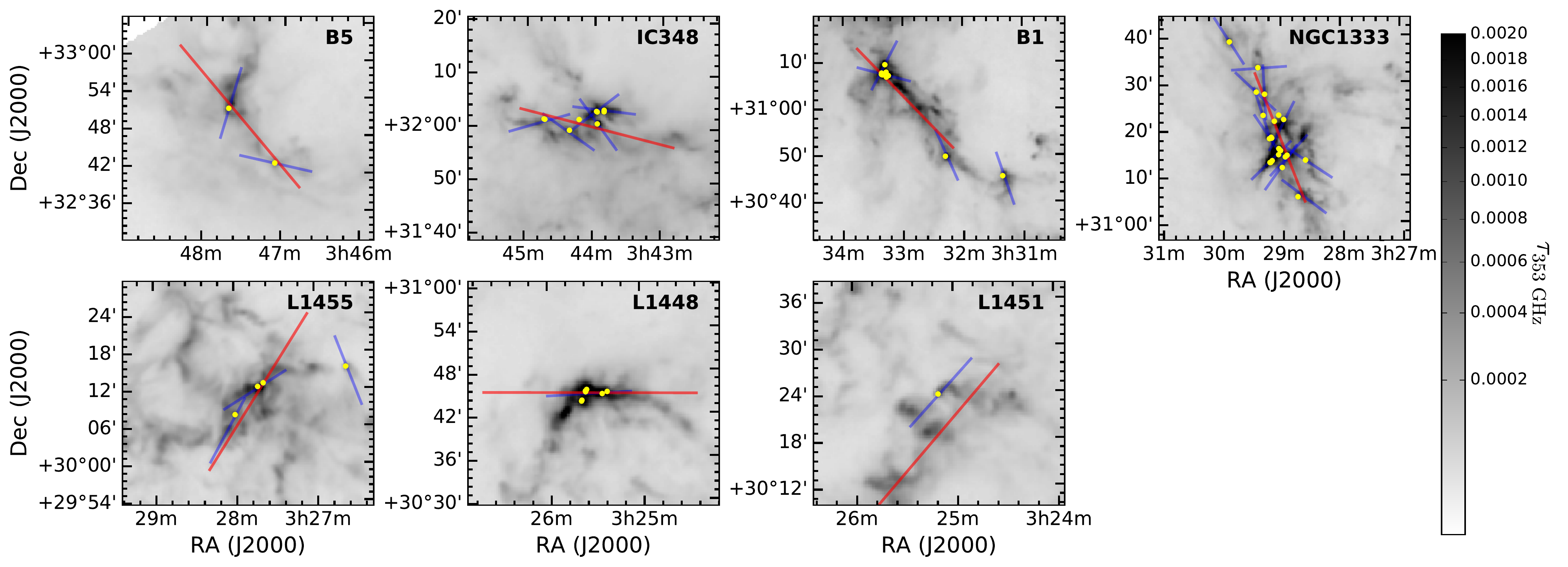}
\end{center}
\caption{\od\ maps \citep{Zari2016} of clumps within the Perseus molecular cloud. Yellow dots show the locations of protostars with measured outflow PAs. The closest blue and red line-centers to each yellow dot represent the small and large scale direction of the filament, respectively, based on fits using SExtractor (essentially a by eye fit; see Section~\ref{filamentPA}). Lines are centered based on the centroid of the SExtractor fit. For both the blue and red lines, the length of the lines are the same angular size in each panel.
}
\label{all_tau_filaments} 
\end{figure*}

\subsection{Outflow Position Angles}\label{outflowPA}
We present the outflow PAs in Table~\ref{tab:angles}. We independently measure the outflow PAs for both the blue- and red-shifted outflows (henceforth, called the blue and red lobes). The range of the PA measurements are from \mbox{--180$^\circ$} to +180$^\circ$; both positive and negative values allow one to assign the appropriate quadrant for the outflow. We also provide the combined PA, \pao, which is simply the average of the two outflows after adding 180$^\circ$ to the lobe with the negative PA. Some entries only provide measurements for one lobe because the other lobe was undetected.

In many cases (about half of the sources) we used outflow PAs from other CO line studies in place of MASSES observations since these studies had data that are better quality and/or at higher resolution. We indicate which study provided the outflow direction for each protostar in the ``Ref/Info" column of Table~\ref{tab:angles}. For the majority of the measured outflow PAs in this study, we made measurements using a methodology very similar to that used in \citet{Hull2013}. We connect the peak intensity of the SMA 1.3\,mm continuum observations with the peak of the integrated intensity maps for both the blue and red outflow lobes. Based on visual inspection, if the CO line emission obviously traces the outflow cavity walls rather than the outflow centroid, we connect the continuum peak to a local CO maximum near the continuum rather than the absolute maximum. In cases where there are no clear local outflow maxima for one lobe, we use the PA measured by the other lobe. If no local maxima exists for both lobes and the CO only traces cavity walls, we manually measure the PA by eye. We indicate in the ``Ref/Info" column of Table~\ref{tab:angles} which outflow measuring method we used. For the angles measured in this paper, a crude approximation of the uncertainty can be found by subtracting the blue outflow PA from the red outflow PA. With such an approximation, the uncertainty in the outflow PA is typically less than 10$^\circ$.

Frequently, the observed field about a MASSES target overlaps with other protostellar sources, which can cause significant confusion in assigning which emission comes from which protostar. To disentangle which emission belongs to which source, we used SAOImage DS9 to overlay all CO emission detected with MASSES on top of \emph{Spitzer} IRAC emission (not shown). In particular, both the 3.6 and 4.5\,$\mu$m $Spitzer$ bands trace the outflow cavities in scattered light and/or knots of H$_2$ emission that are most prominent in the 4.5\,$\mu$m channel. We also use the catalog of Perseus protostars from \citet{Young2015} to locate other nearby T Tauri stars that may be contributing to the CO emission observed by the SMA. Together, we are able to disentangle which outflow emanates from which source. In this paper, we only present the outflow PAs that we believe we were confidently able to determine. Protostars surveyed by MASSES that are not presented in this paper were either not yet imaged or had confusing CO emission that did not allow for a reliable measurement of \pao. In total, we have \pao\ measurements for 57 protostellar outflows. In Figure~\ref{all_tau_outflows} we overlay each \pao\ measurement on the $Herschel$-derived \od\ map. The SMA CO(2--1) integrated intensity maps for two protostars are shown in the right panels of Figure~\ref{filament_fit}; other sources can be found in the electronic version of the paper. The average spectra within the vicinity of the protostar (i.e., within a radius of 8$\arcsec$) is shown in Figure~\ref{spectra}.




\subsection{Filament Direction}\label{filamentPA}
We present the filament PAs in Table~\ref{tab:angles}. We determine the filament directions based on $Herschel$-derived \od\ maps (see Section~\ref{opticaldepth}). Since extracting filaments directions can sometimes depend on the method used, we use two different techniques. One technique is based on \texttt{FILFINDER} and the other is based on SExtractor. For both techniques, we also investigate how the filament directions depend on both small and large scale optical depth characteristics.

\subsubsection{Using \texttt{FILFINDER} for Filament Position Angles}

The first method extracts the filamentary structure using the \texttt{FILFINDER} algorithm \citep{Koch2015} as implemented in \texttt{PYTHON}. \texttt{FILFINDER} is unique in that it can find filaments with relatively low surface brightness compared to the main filaments, which is achieved by using an arctangent transform on the image. This algorithm first isolates the filamentary structure across the entire map. Then, each filament within the filamentary structure is made into a one-pixel-wide skeleton via the Medial Axis Transform \citep{Blum1967}. We use the default implemented parameters in the \texttt{FILFINDER} algorithm, with the exception of the parameters \texttt{size\_thresh} and \texttt{skel\_thresh}, which were altered to provide the best visual fit to the actual Perseus data. Specifically, for these parameters we used the values \texttt{size\_thresh}~=~300 and \texttt{skel\_thresh}~=~100. The resolution of the observations (36$\arcsec$) and the distance to the Perseus molecular cloud (235\,pc) were also provided to the \texttt{FILFINDER} algorithm.

\texttt{FILFINDER} determines the filament direction via the Rolling Hough Transform \citep{Clark2014}. Unfortunately, the Rolling Hough Transform often performs poorly in the Perseus molecular cloud since \texttt{FILFINDER} sometimes combines distinct molecular clumps as a single filamentary structure. For example, \texttt{FILFINDER} combines NGC~1333 and L1455 into a single filamentary network and measures the direction of the combined structure. We find that in most of these instances, the Rolling Hough Transform poorly estimates both the small and large scale filamentary direction. Instead of this transform, we approximate the filamentary direction by fitting a line to the filamentary skeleton output from \texttt{FILFINDER}. To do this, we first find the closest \texttt{FILFINDER} skeleton pixel to the position of the protostar given by \citet{Tobin2016}. We then extract a square skeleton map of 11$\times$11~pixels (198$\arcsec$~$\times$~198$\arcsec$ or $\sim$0.2\,pc~$\times$~0.2\,pc) centered on this closest skeleton pixel and fit an ordinary least squares bisector line \citep{Isobe1990,Feigelson1992} to the scatter plot of the skeleton pixels. The slope of this fitted line is then converted to a PA. We use an extraction of an 11$\times$11~pixel square because we find it large enough to fit the elongation of the filament, but small enough that the filament's direction is not strongly influenced by other nearby filamentary structures. We have also ran the same algorithm for extracting squares of skeleton pixels that are up to $\sim$3 times larger or smaller than 11$\times$11~pixels, and the results in our paper are qualitatively the same. The 11$\times$11~pixel extraction provides the best visual fits to the filaments across all sources.

Figure~\ref{filament_fit} shows examples of this fitting process for two sources; other sources can be found in the electronic version of this paper. Note that the measured PA (red line in middle panels of Figure~\ref{filament_fit}) are slightly off as one may measure by eye simply because nearby filament branches in the 11$\times$11~pixel cutout of the skeleton map affects the bisector fit. In the rest of the paper, we will refer to this method for extracting filament directions as the ``\texttt{FILFINDER} algorithm." In Table~\ref{tab:angles}, we provide these filament angles, \pafilf, along with their corresponding projected outflow-filament angle, \gaf.

Angular momentum of a protostar could possibly be inherited from filamentary structures larger than the filaments measured with 36$\arcsec$ resolution. Therefore, we also make a comparison to larger scales by Gaussian smoothing the \citet{Zari2016} \od\ maps and rerunning the \texttt{FILFINDER} algorithm discussed above. Specifically, we smooth the data to resolutions of 1$\arcmin$, 2$\arcmin$, 3$\arcmin$, 4$\arcmin$, 5$\arcmin$, and 6$\arcmin$, where 1$\arcmin$ is 0.068\,pc, assuming a distance of 235\,pc to Perseus. \texttt{FILFINDER} progressively finds fewer branches in the Perseus filaments when we smooth \od\ maps to these coarser resolutions. The measured projected outflow-filament angles for these resolutions are shown in Table~\ref{tab:angles} as $\gamma_{\rm{X}\arcmin}$, where X\arcmin\ is the smoothed resolution in arcminutes.

\subsubsection{Using SExtractor for Filament Position Angles}
The second method fits ellipses to the filaments via SExtractor \citep{Bertin1996}, as implemented in the Graphical Astronomy and Image Analysis (GAIA) Tool\footnote{\url{http://star-www.dur.ac.uk/~pdraper/gaia/gaia.html}}. 
SExtractor works by fitting ellipses to the emission data.  We then adopt the position angle of the fitted ellipses as the filament PA. To measure both the large scale and small scale filamentary structure, we extract two different filament directions for each protostar. For the large scale structure, we fit a single filamentary direction to the clump (i.e., the pc-scale cloud structure), and for the small scale, we fit the most localized elongated structure for the protostar. For both scales, the parameters \texttt{Detection threshold}, \texttt{Analysis threshold}, and \texttt{Contrast parameter} were adjusted for each source so that the fitted ellipse best matches the elongation as judged by the human eye. We find that no single set of values for these three parameters can fit all filaments in the Perseus cloud that is agreeable with the human eye, and thus the parameters were adjusted filament-by-filament. Therefore, this method is primarily a ``by eye" determination of the filament direction with the aid of software. This method of determining the filament PA is very similar to the method used in \citet{Anathpindika2008}. We note that even at the small scale, the best SExtractor fit for a local filament may be the same for multiple protostars. 

Figure~\ref{all_tau_filaments} shows both the small and large scale filament PAs determined for each protostar using this method. The final projected outflow-filament angles using this method for both the small scale ($\gamma_{\mbox{se,S}}$) and large scale ($\gamma_{\mbox{se,L}}$) are given in Table~\ref{tab:angles}. The measured filament PAs for both of these methods can be derived from $\gamma_{\mbox{se,S}}$ and $\gamma_{\mbox{se,L}}$ by using Equation~\ref{filequation} and the individual \pao\ measurements.

\subsubsection{Comparison of the \texttt{FILFINDER} and SExtractor Techniques}
\begin{figure}[ht!]
\begin{center}
\includegraphics[width=1\columnwidth]{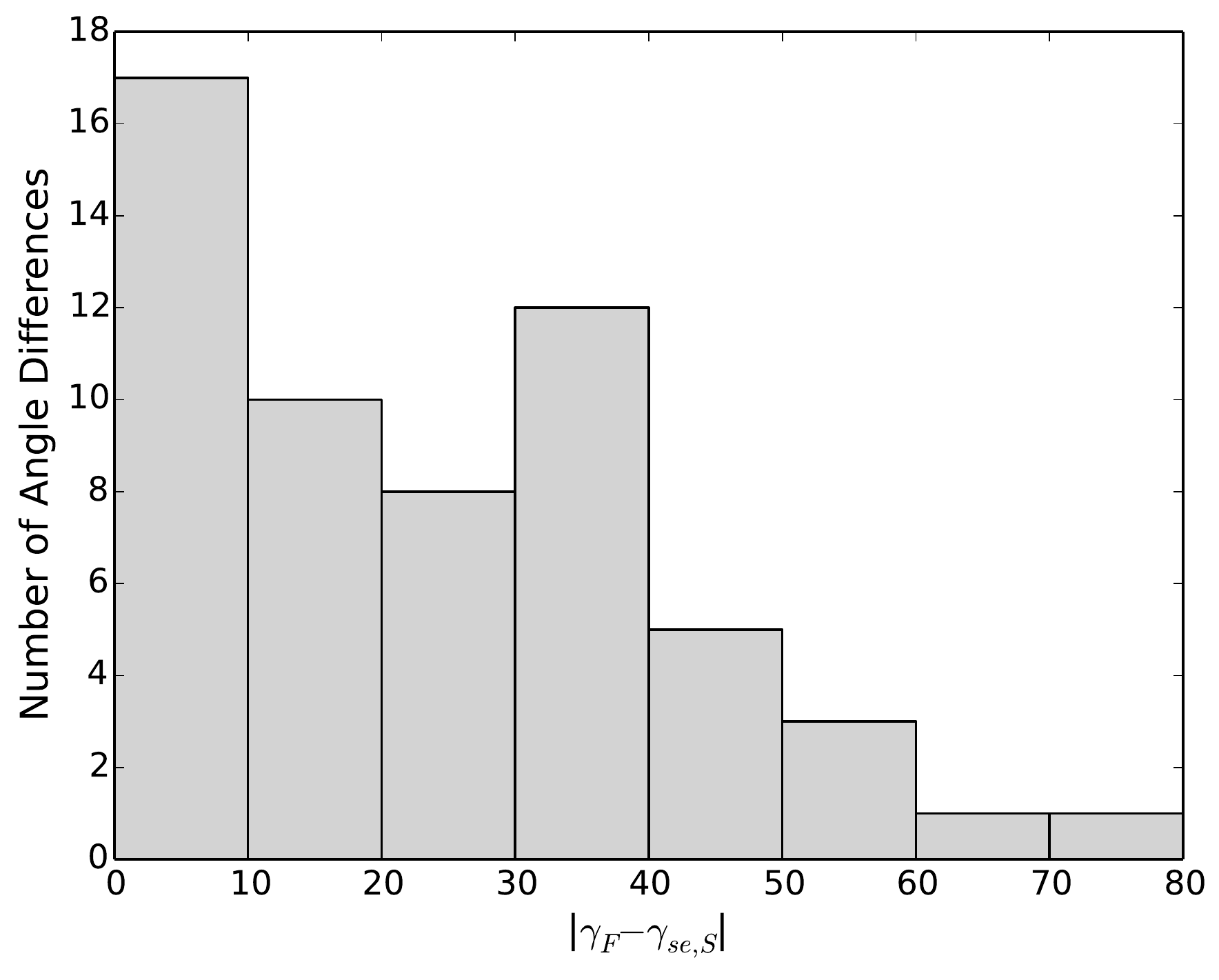}
\end{center}
\caption{
Histogram showing the magnitude of the difference in the projected outflow-filament angles measured by the two methods used to find the filament orientation. $\gamma$ values for the \texttt{FILFINDER} algorithm and the small scale SExtractor fits are indicated by \gaf\ and $\gamma_{\mbox{se,S}}$, respectively.
}
\label{gamma_difs} 
\end{figure}

Both the \texttt{FILFINDER} and SExtractor filament-finding methods have their advantages and disadvantages. For example, the first method is completely automated, and if there are multiple filamentary branches in the field, the algorithm attempts to find the best filamentary direction in a fixed area of $\sim$0.2\,pc~$\times$~0.2\,pc. However, filamentary branches may be considered as a contaminate, in which case the second method (the SExtractor by eye measurement) may more accurately determine the filamentary direction.

When comparing the two methods, the filament direction found with the \texttt{FILFINDER} algorithm are most comparable to those found at small scale with SExtractor since these both measure filaments at approximately the same size scales. Figure~\ref{gamma_difs} shows the absolute value of the difference in the measured angles \gaf\ and $\gamma_{\mbox{se,S}}$ for each protostar. Since \pao\ for each protostar is measured the same regardless of the method used to measure the filament orientation, \gaf~--~$\gamma_{\mbox{se,S}}$ is equivalent to the difference in the measured filament directions for each technique. This histogram shows that the measured filament position angles mostly agree, but in some cases, the measured filament angles for each technique vary significantly. Therefore in the following section, we present statistical comparisons to the outflows for both filament-finding techniques.



\begin{figure}[ht!]
\begin{center}
\includegraphics[width=1\columnwidth]{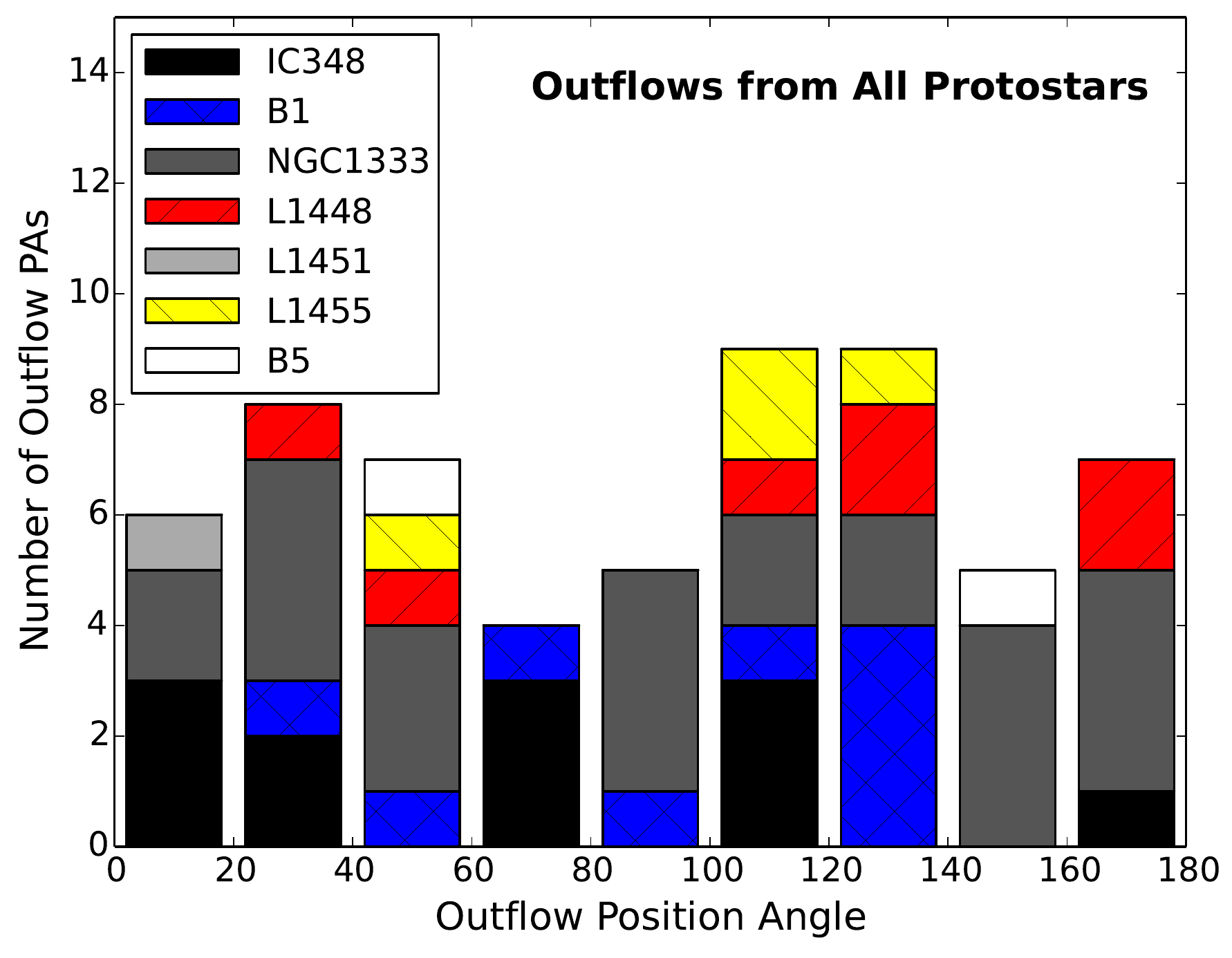}
\end{center}
\caption{Stacked histogram (with 20$^\circ$ bins) of outflow PAs in the Perseus molecular cloud. Colors correspond to the clump that \pao\ belongs to.
}
\label{outflow_histo} 
\end{figure}

\begin{figure}[ht!]
\begin{center}
\includegraphics[width=1\columnwidth]{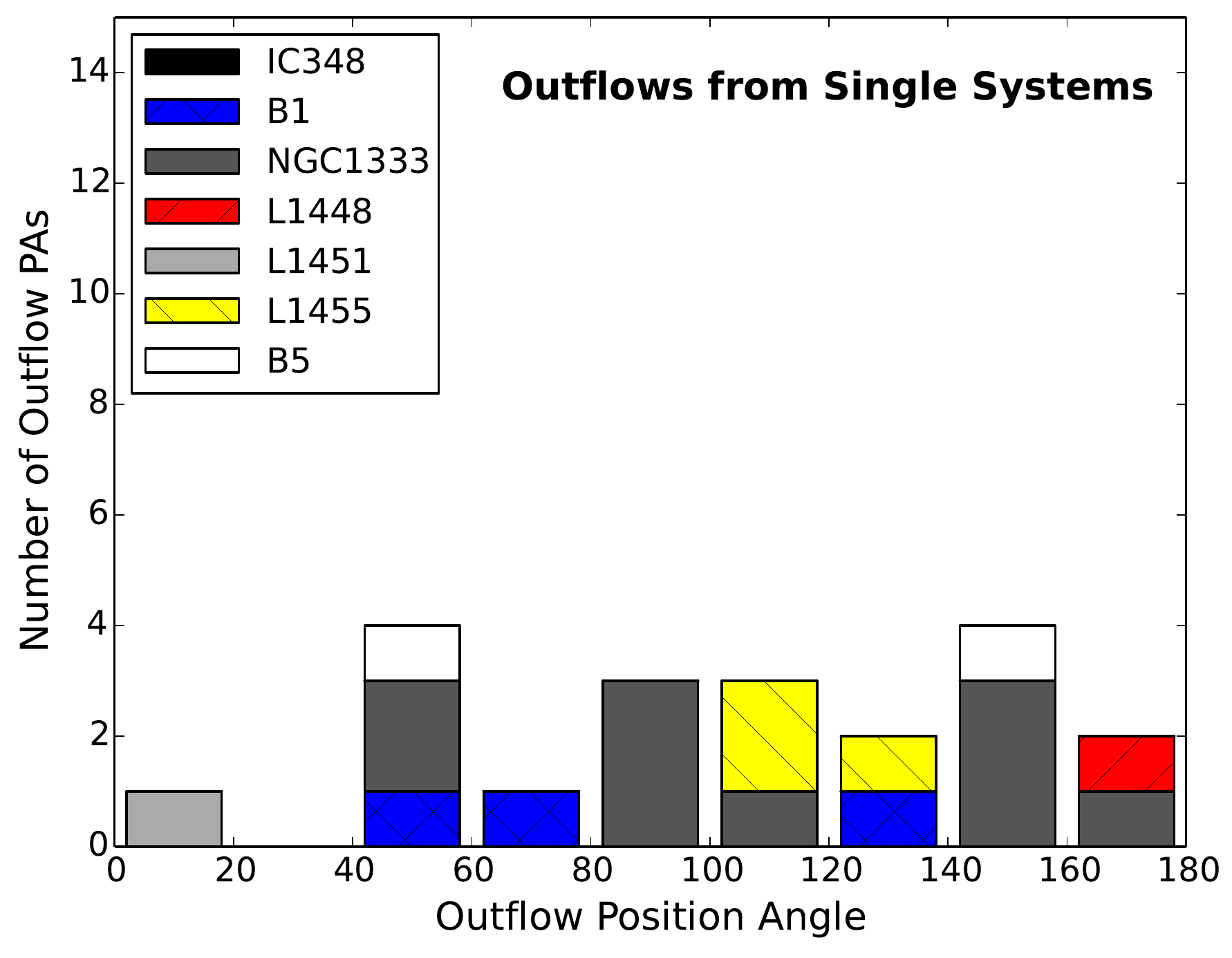}
\end{center}
\caption{Same as Figure~\ref{outflow_histo} but now only considering protostars that were not identified as multiples in the VANDAM survey \citep{Tobin2016}.
}
\label{outflow_histo_singles} 
\end{figure}

\section{Results}\label{results}
In this section, we analyze the distributions of \pao\ and $\gamma$. We note that, when we compare our empirical distributions of angles \pao\ and $\gamma$ to simulated data, we favor the Anderson--Darling (AD) test \citep[e.g.,][]{Stephens1974} over the Kolmogorov--Smirnov test. The AD test tends to be more powerful in detecting differences in distributions than the Kolmogorov--Smirnov test, particularly at the tail ends of the distributions \citep[e.g.,][]{Hou2009,Engmann2011,Razali2011}. While the $p$-values differ for these two tests, the overall statistical significance does not change dramatically and our conclusions remain unchanged. For the two-sample AD test, $p$-values near 1 imply that the two distributions are likely drawn from the same distribution, while $p$-values near 0 imply that they are unlikely drawn from the same distribution.

\subsection{Outflow Directions in Perseus}
Figure~\ref{outflow_histo} shows a stacked histogram of \pao, where the color of each stacked bar indicates the protostar's parental clump. As with Figure~\ref{all_tau_outflows}, this figure does not show any obvious relationship between \pao\ and the protostar's parental clump. Since a stellar companion could possibly affect the spin axis of a protostar \citep[e.g.,][]{Offner2016,LeeJoyce2017}, we also show a stacked histogram of the ``single" systems identified in the VANDAM survey \citep{Tobin2016} in Figure~\ref{outflow_histo_singles}. This survey used multi-wavelength data with resolutions as high as 15\,AU, and defined a system as a ``single" system if it has no detected companions within 10,000\,AU. Again, the distribution is mostly random. We compare the ``all" and ``single system" data to a random distribution, and the AD test gives $p$-values of 0.65 and 0.62, respectively. This signifies that we cannot distinguish the \pao\ histograms in Figures \ref{outflow_histo} and \ref{outflow_histo_singles} from a random distribution of angles.

\begin{figure}[ht!]
\begin{center}
\includegraphics[width=1\columnwidth]{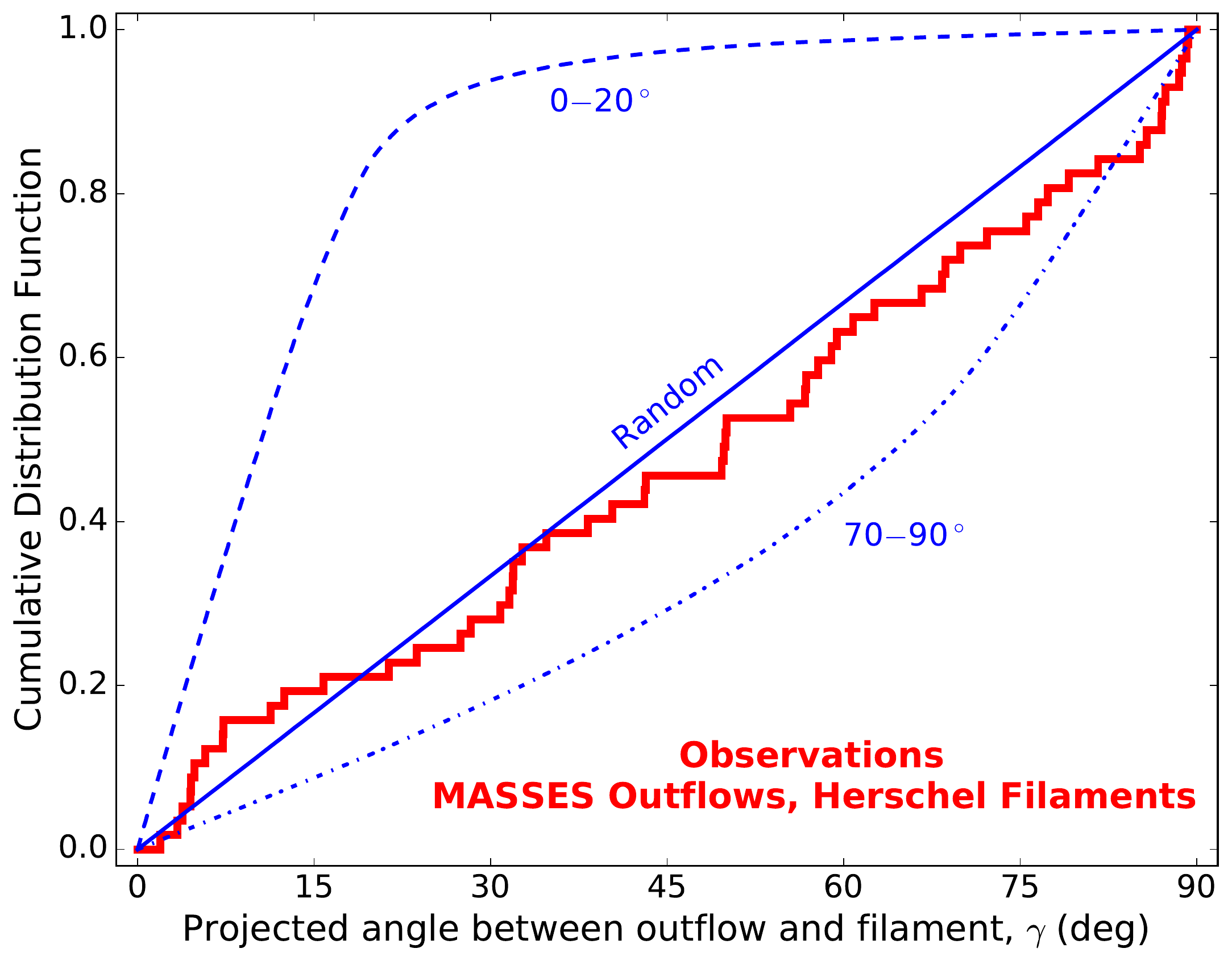}
\end{center}
\caption{Cumulative distribution function of the projected angles between outflows and filaments, $\gamma$. The red step function shows \gaf\ for this study, which measures the angle between MASSES outflows and fitted $Herschel$ filaments directions using the \texttt{FILFINDER} algorithm discussed in Section~\ref{filamentPA}. The three blue lines show Monte Carlo simulations of the expected projected $\gamma$ for outflows and filaments that are 3-dimensionally only parallel (actual outflow-filament angle that is between 0 and 20$^\circ$), only perpendicular (70--90$^\circ$), or completely random (0--90$^\circ$).
}
\label{main_cdf} 
\end{figure}

\subsection{Cumulative Distribution Functions using FILFINDER Filament Angles}\label{cdf_filfinder}
While the first visual and clump regional tests did not show any obvious relationship between clump structure and protostellar outflow directions, clumps are pc-sized while filaments are about 0.1\,pc in diameter \citep[e.g.,][]{Arzoumanian2011}. As discussed in Section~\ref{filamentPA}, we use \texttt{FILFINDER} to extract filament directions at the 36$\arcsec$ (0.04\,pc) scale. These filament directions, \pafilf, are then compared to \pao\ to determine the projected outflow and filament angular difference, \gaf. We plot the cumulative distribution function (CDF) of the observed \gaf\ in Figure~\ref{main_cdf}. To investigate whether the distribution of \gaf\ reflects outflows and filaments that are primarily aligned parallel, perpendicular, or at random, we perform 3-D Monte Carlo simulations that we project onto 2-D. Specifically, we simulate the CDF of the expected projected angles in the sky for outflow-filament angles that are 3-dimensionally ``only parallel" (defined as actual outflow-filament angles that are distributed between 0$^\circ$ and 20$^\circ$), ``only perpendicular" (actual angles between 70$^\circ$ and 90$^\circ$), or completely random (actual angles between 0$^\circ$ and 90$^\circ$). The expected observed (i.e., projected) $\gamma$ for these three Monte Carlo instances are also shown in Figure~\ref{main_cdf}. Detailed information on the Monte Carlo simulations is presented in Appendix~\ref{montecarlo}.

\begin{figure}[ht!]
\begin{center}
\includegraphics[width=1\columnwidth]{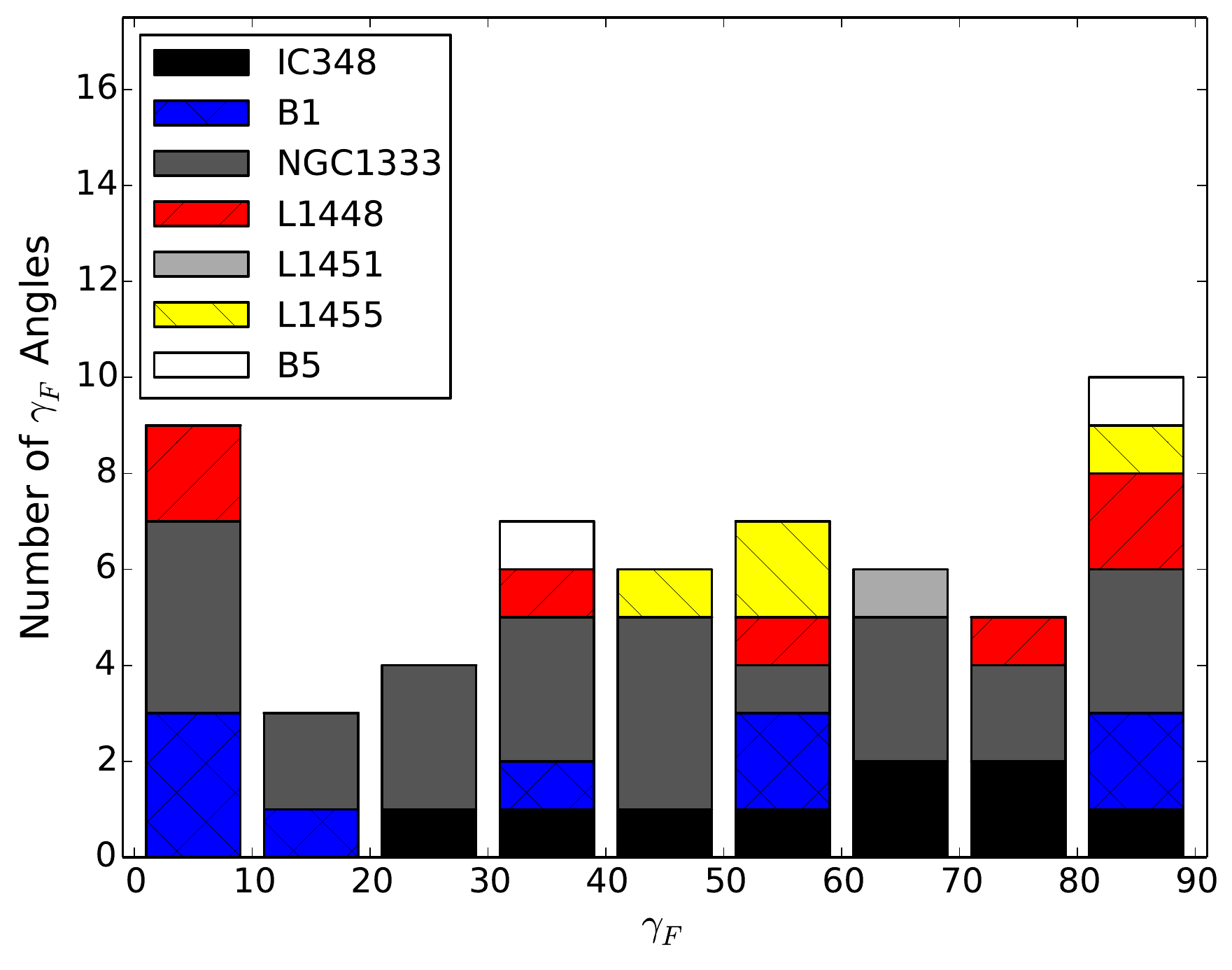}
\end{center}
\caption{Same as Figure~\ref{outflow_histo}, but now the stacked histogram is shown for \gaf. The histogram bin size is 10$^\circ$.
}
\label{gamma_histo} 
\end{figure}

Immediately evident from Figure~\ref{main_cdf} is that the distribution of \gaf\ is inconsistent with outflows and filaments that are preferentially parallel. The projected angles are also inconsistent with a purely perpendicular alignment with over 99\% confidence (AD test gives a p-value = 0.0045). However, we cannot significantly distinguish the \gaf\ distribution from a distribution of randomly aligned outflows and filaments ($p$-value~=~0.20). Table \ref{tab:pvals} summarizes the statistical tests conducted on all the $\gamma$ measurements discussed in Section~\ref{dataanalysis}. In Figure~\ref{gamma_histo}, we show the distribution of \gaf\ as a stacked histogram, with colors representing the parental clump. No obvious non-random relationship is found, regardless of the protostar's clump location.

\begin{figure}[ht!]
\begin{center}
\includegraphics[width=1\columnwidth]{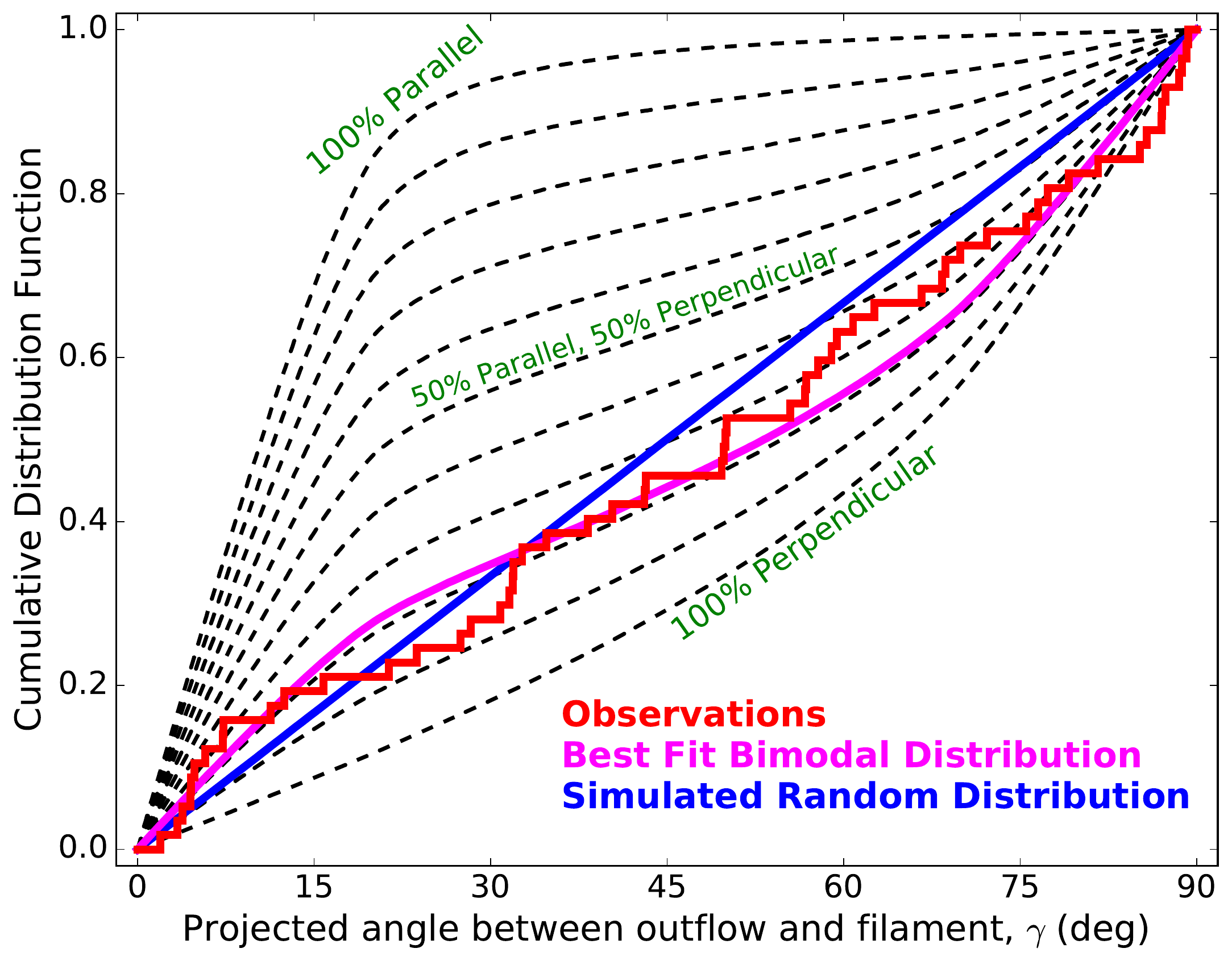}
\end{center}
\caption{Cumulative distribution function of the projected angles between outflows and filaments, $\gamma$, with the red step curve showing the empirical distribution, \gaf. Black dashed lines show different mixes of projected outflow-filament angles that are 3-dimensionally parallel and perpendicular in increments of 10\% (i.e., the top line is 100\% parallel and 0\% perpendicular, the next line is 90\% parallel and 10\% perpendicular, and so on). Parallel angles are defined as 3-dimensional angles drawn from a distribution between 0$^\circ$ and 20$^\circ$, while perpendicular angles are defined as angles drawn from a distribution between 70$^\circ$ and 90$^\circ$ (see Appendix~\ref{montecarlo} for details). The blue line shows a random distribution of projected angles, while the magenta line shows the best bimodal fit to the data of 22\% parallel and 78\% perpendicular.
}
\label{bimodal_cdf} 
\end{figure}

So far, we produced simple models of $\gamma$ from outflow-filament angles that are only parallel, only perpendicular, or aligned at random. As mentioned in Section~\ref{sec:intro}, outflow orientation may be determined by the dominant flow direction about the filament. Therefore, a bimodal distribution of $\gamma$ is possible, e.g., a mix of both parallel and perpendicular orientations. 

We test different 3-dimenisonal combinations of purely parallel (again, where angles are distributed between 0$^\circ$ and 20$^\circ$) and purely perpendicular (angles between 70$^\circ$ and 90$^\circ$) outflow-filament angles via Monte Carlo simulations.  We consider 101 bimodal cases in increments of 1\% (i.e., 100\% parallel, 99\% parallel and 1\% perpendicular, 98\% parallel and 2\% perpendicular, ..., 100\% perpendicular). Figure~\ref{bimodal_cdf} shows the CDFs of several of these bimodal distributions projected into 2 dimensions. We find that, when comparing to the observed distribution \gaf, the simulated $\gamma$ that is a mix of 22\% parallel and 78\% perpendicular maximizes the $p$-values for the AD test (as well as the Kolmogorov--Smirnov test). The $p$-value for this case is 0.55, signifying a slightly more consistent distribution with the observed \gaf\ distribution than a random distribution.

This bimodal test can also constrain which mixes of parallel and perpendicular are unlikely. According to the AD test, we find that at 95\% confidence, \gaf\ probably does not come from a bimodal distribution that is more than 39\% parallel or more than 94\% perpendicular. At 85\% confidence, we find that the \gaf\ distribution does not come from a bimodal distribution that is more than 33\% parallel or more than 90\% perpendicular.

Other mixes of $\gamma$ distributions are also possible, such as mixes of a random distribution with perpendicular and/or parallel distributions. We do not test other distribution mixes in this paper since we mainly want to show that perpendicular outflows and filaments are much more likely than parallel.

\begin{figure}[ht!]
\begin{center}
\includegraphics[width=1\columnwidth]{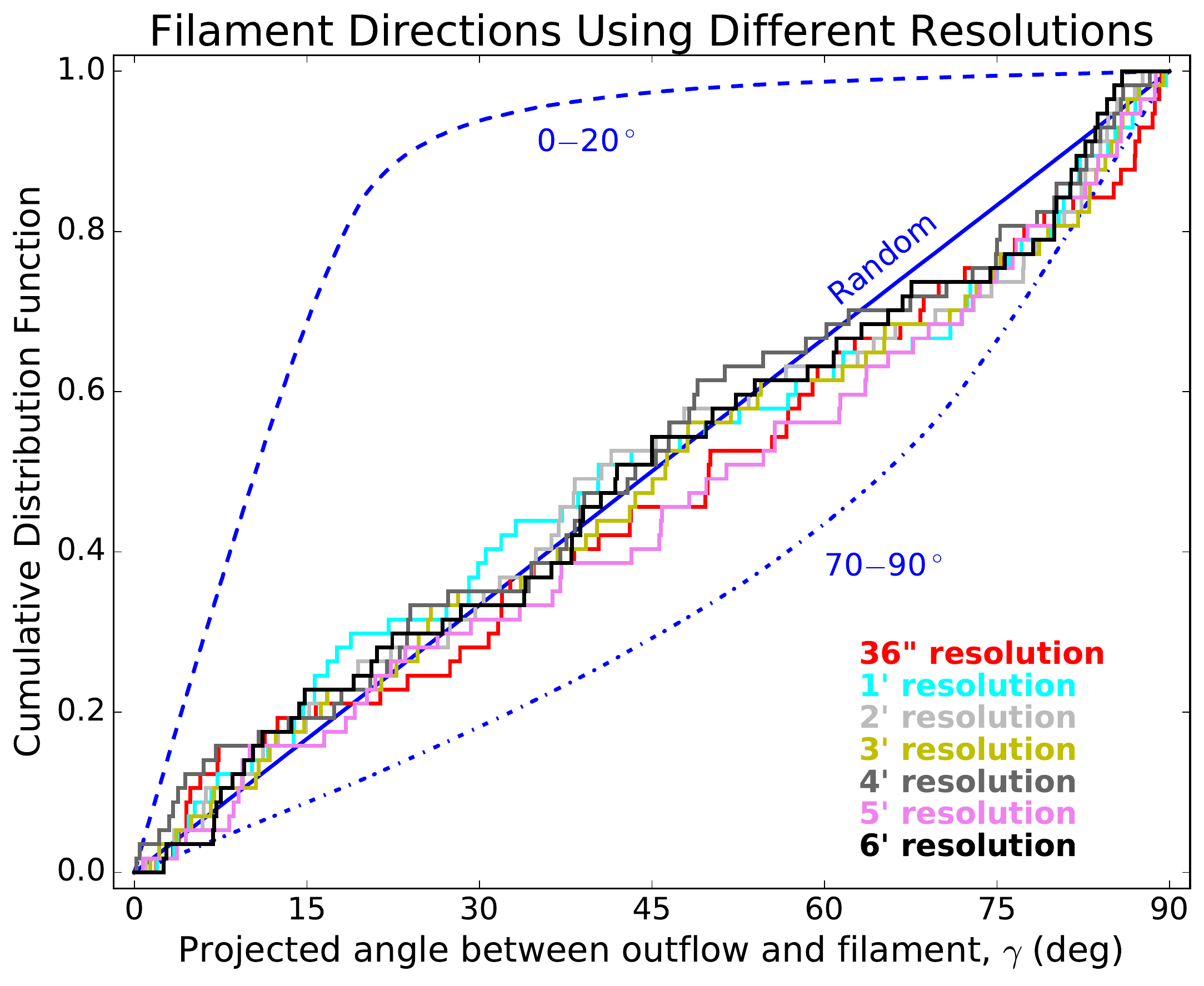}
\end{center}
\caption{Figure caption is the same as Figure~\ref{main_cdf}, except with additional step curves showing the effects of smoothing the Perseus \od\ map before running the \texttt{FILFINDER} algorithm described in Section~\ref{filamentPA}. The colors indicate which resolution the \od\ map was smoothed to in creating the empirical $\gamma$ CDF.
}
\label{multiscale} 
\end{figure}



\begin{deluxetable}{l@{}c@{}c@{}}
\tablecolumns{4}
\tabletypesize{\tiny}
\tablewidth{0pt}
\tablecaption{Anderson--Darling Test $p$-values \label{tab:pvals}}
\tablehead{\colhead{Empirical $\gamma$} & \colhead{$p$-value, compared} & \colhead{$p$-value, compared} \\
\colhead{Distribution} & \colhead{with Random} & \colhead{with Perpendicular}
}
\startdata
\gaf & 0.20 & 0.0045 \\
\pafone & 0.33 & 0.00085 \\
\paftwo & 0.40 & 0.0011 \\
\pafthree & 0.42 & 0.0029 \\
\paffour & 0.49 & 0.00024 \\
\paffive & 0.24 & 0.023 \\
\pafsix & 0.59 & 0.00069 \\
$\gamma_{\mbox{se,S}}$ & 0.74 & 0.00014 \\
$\gamma_{\mbox{se,L}}$ & 0.64 & 0.0021 \\
Anathpindika \& Whitworth  & 0.017 & 0.16 \\
\gaf, Single Protostars & 0.60 & 0.18 \\
\gaf, Multiple Protostars & 0.20 & 0.011 \\
\gaf, with $T_{\rm{bol}} < 50$ & 0.53 & 0.021 \\
\gaf, with $T_{\rm{bol}} > 50$ & 0.18 & 0.075 \\
\gaf, with \od\ $<$ 0.016 & 0.15 & 0.20 \\
\gaf, with \od\ $>$ 0.016 & 0.27 & 0.0050
\enddata
\tablecomments{$p$-values are not shown for empirical $\gamma$ distributions compared with parallel $\gamma$ distributions because they are all extremely low in value (less than 10$^{-9}$).}
\end{deluxetable}

As discussed in Section~\ref{filamentPA}, we also determine filament angles by running the \texttt{FILFINDER} algorithm on Perseus \od\ maps that have been smoothed to coarser resolution. The resulting CDFs for $\gamma$ at these resolutions are shown in Figure~\ref{multiscale}. We find that the CDFs at all resolutions are similar with each other, with the AD test $p$-value 0.45 or greater when comparing any two distributions. We also find consistent results between the smoothed and the non-smoothed (36$\arcsec$) resolution $\gamma$ angles. Specifically, as shown in Table \ref{tab:pvals}, none of the $\gamma$ distributions extracted from the smoothed \od\ maps can be statistically distinguished from a random distribution, but all are inconsistent with projected angles from an only perpendicular and only parallel distributions.

\begin{figure}[ht!]
\begin{center}
\includegraphics[width=1\columnwidth]{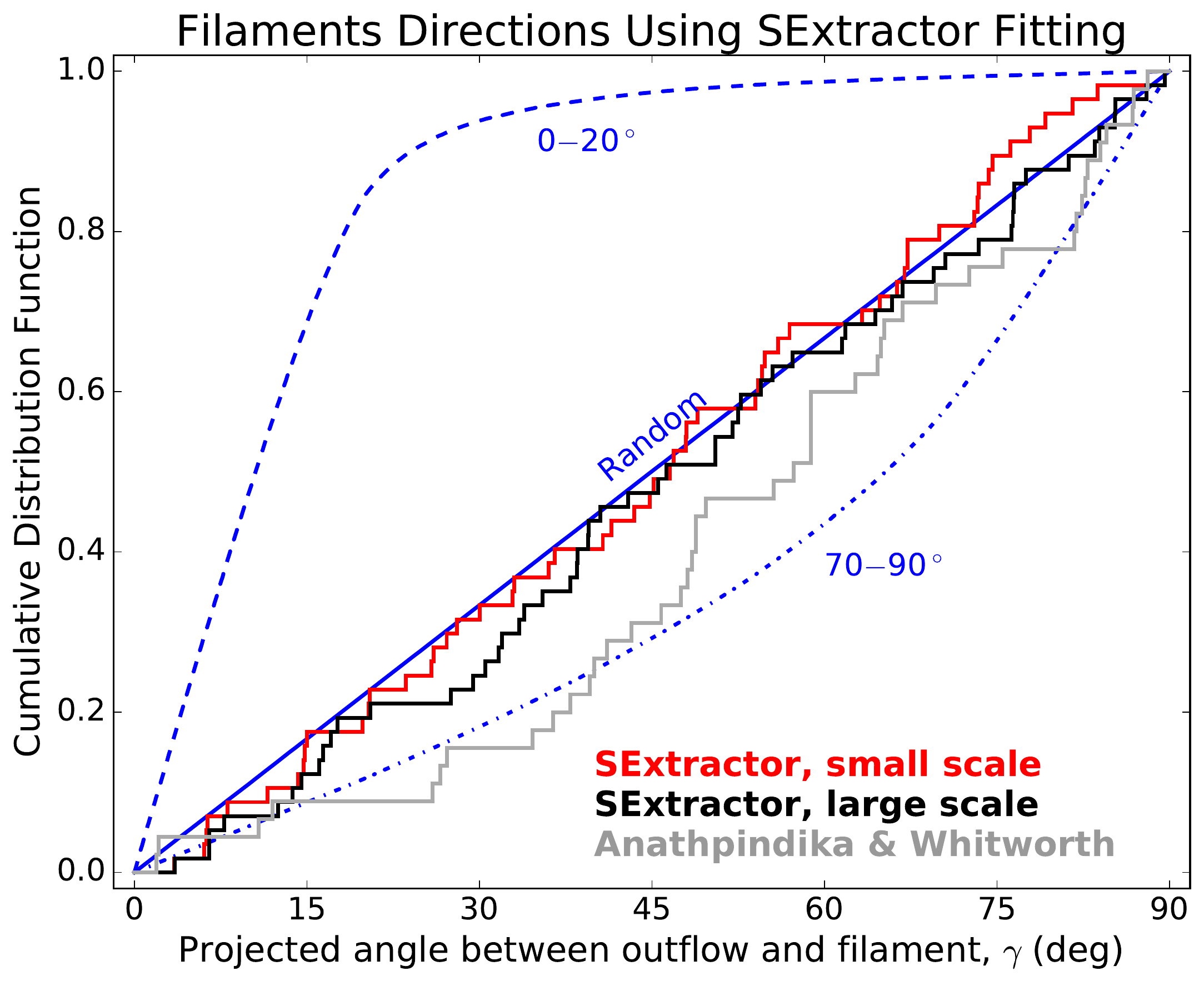}
\end{center}
\caption{Figure caption is the same as Figure~\ref{main_cdf}, except now the step functions show the SExtractor fitting of filament directions (which is essentially a ``by eye" fit) at both small and large scales, as described in Section~\ref{filamentPA}. The results from \citet{Anathpindika2008} are also shown, but we caution any interpretation of this curve due to shortcomings of the study discussed in Appendix~\ref{anathdiscuss}.
}
\label{sextractor} 
\end{figure}


\subsection{Cumulative Distribution Functions using SExtractor Filament Angles}
In Figure~\ref{sextractor}, we show the CDFs when using the SExtractor filament direction fits, which is essentially a fit by eye (see Section~\ref{filamentPA}). We find similar results for both the small-scale (i.e., fitting the closest elongated feature to each protostar) and large-scale (i.e., fitting the main part of the clump containing each group of protostars) SExtractor fitting as with the \texttt{FILFINDER} algorithm.  That is, the SExtractor fits are not inconsistent with a random distribution and are significantly inconsistent with both parallel and perpendicular angle distributions (see Table \ref{tab:pvals}). Also shown in this figure are the results from \citet{Anathpindika2008}, which uses a similar filament fitting algorithm. Unlike our results, their distribution for $\gamma$ is more consistent with perpendicular ($p$-value of 0.17) than random ($p$-value of 0.017). However, we caution an interpretation of the \citet{Anathpindika2008} $\gamma$ distribution due to several shortcomings in their study, which are discussed in detail in Appendix~\ref{anathdiscuss}. We do not show the results from \citet{Davis2009} because they do not supply any information on $\gamma$ or the filament PAs.

As in Section~\ref{cdf_filfinder}, we also test which bimodal distribution of parallel and perpendicular projected orientations matches the observations using SExtractor filament fits. The results are similar as those found with \texttt{FILFINDER}.


\begin{figure*}[ht!]
\begin{center}
\includegraphics[width=0.67\columnwidth]{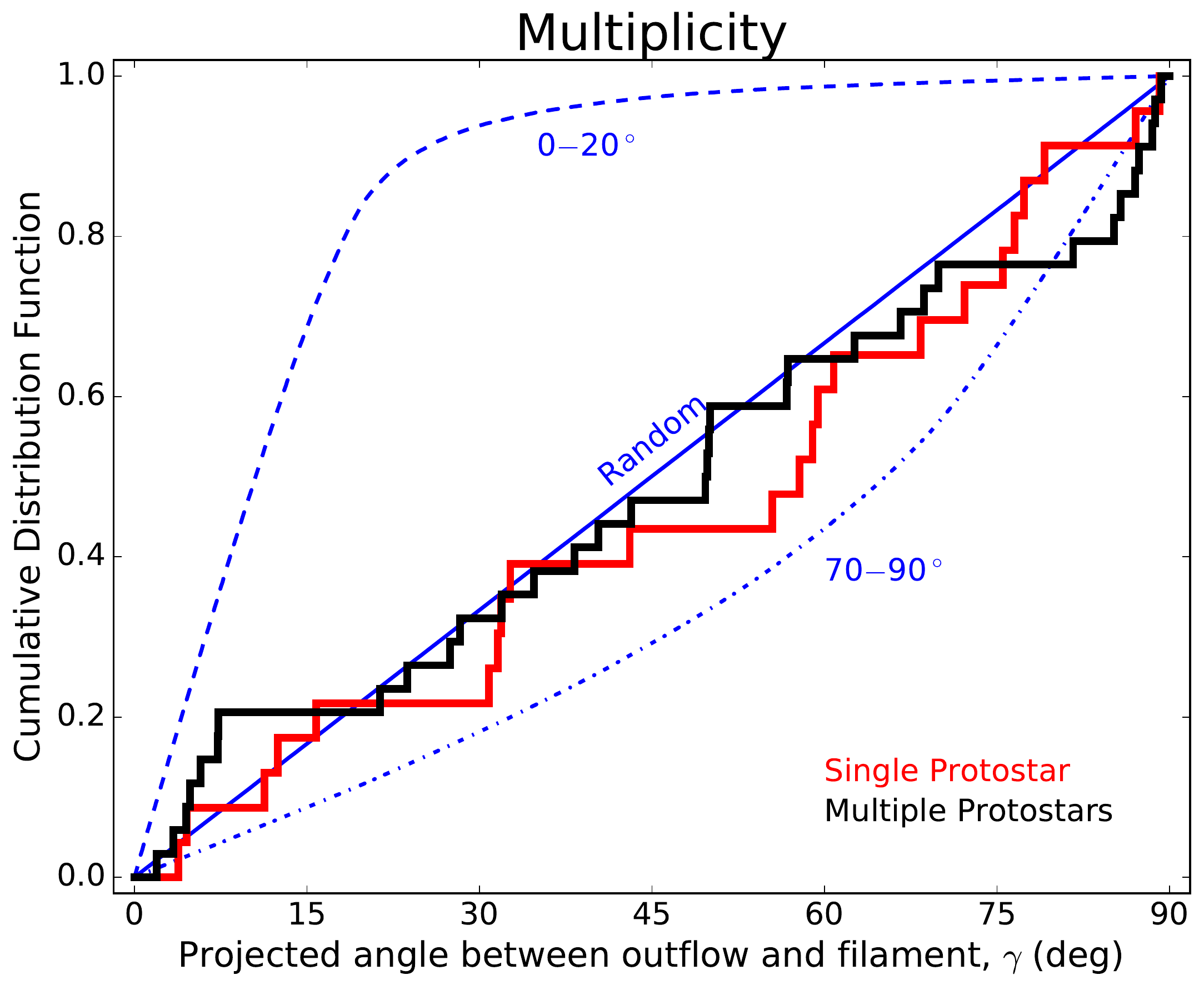}~~~
\includegraphics[width=0.67\columnwidth]{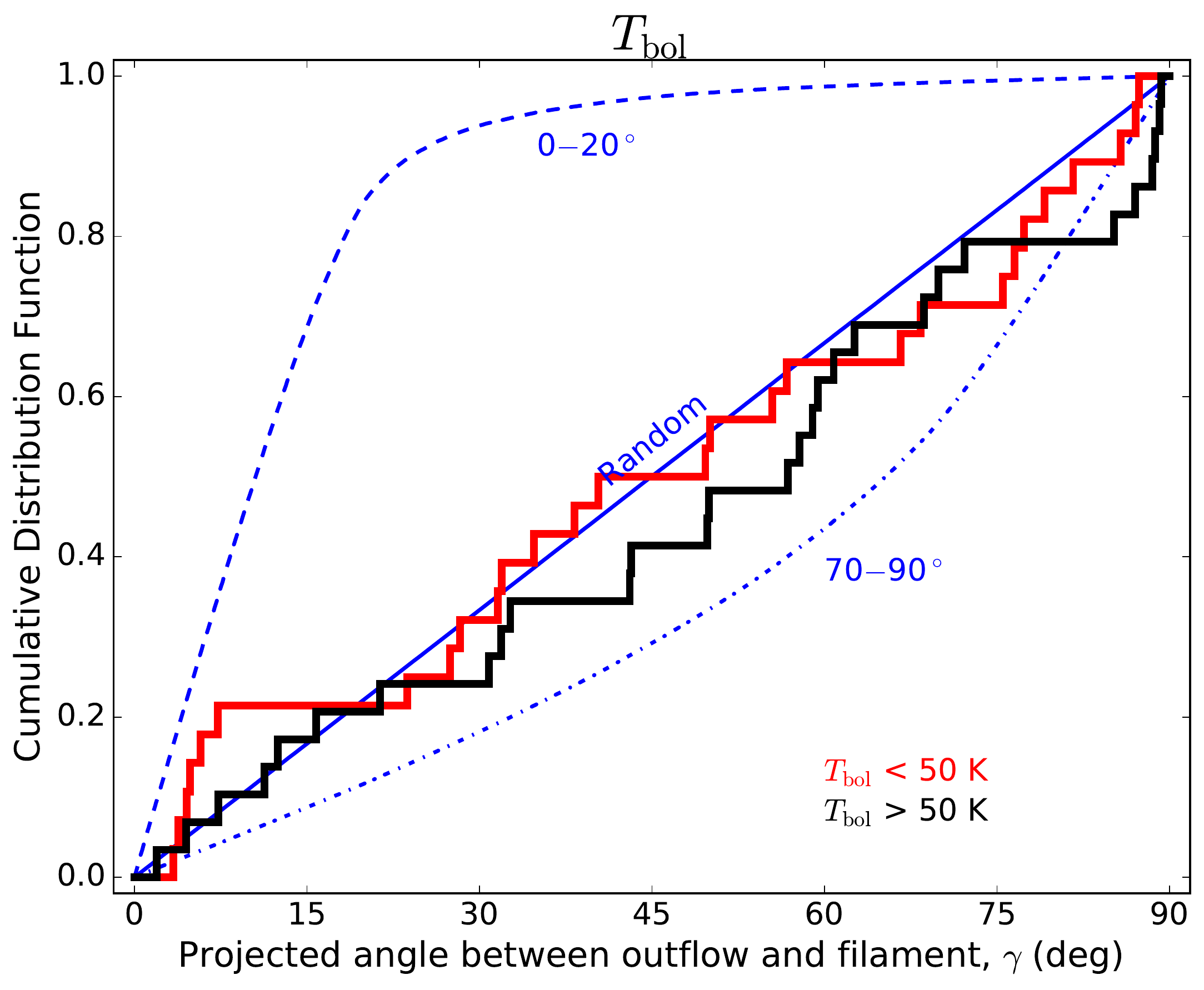}~~~
\includegraphics[width=0.67\columnwidth]{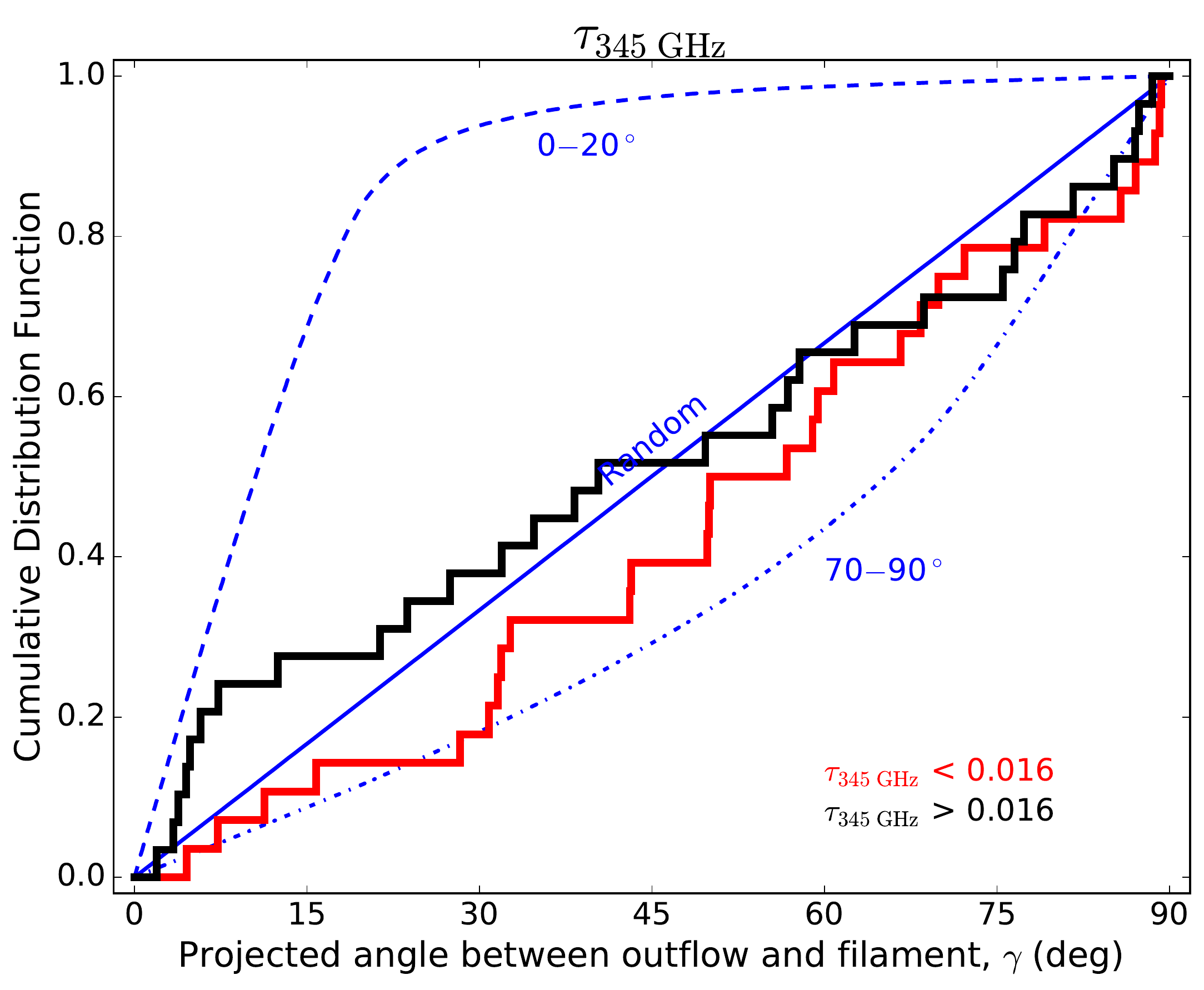}
\end{center}
\caption{CDFs of $\gamma$, binning data based on multiplicity (left), bolometric temperature (middle, an indicator of age), and optical depth (right). All CDFs use filament measurements from the \texttt{FILFINDER} algorithm.
}
\label{knobs} 
\end{figure*}

\subsection{Cumulative Distribution Functions Based on Protostellar Characteristics}
Here we investigate whether or not individual characteristics of the protostars themselves or their surrounding environment affect the underlying $\gamma$ distribution. We consider the protostar's multiplicity, the protostar's bolometric temperature ($T_{\rm{bol}}$), and the \citet{Zari2016} \od\ pixel value at the protostar. Both the protostellar multiplicity and $T_{\rm{bol}}$ were taken from \citet{Tobin2016} and references therein. For multiples that were resolved with the VLA but not with $Spitzer$, we assign the same $T_{\rm{bol}}$ for all multiples within the $Spitzer$-defined source. The left panel of Figure~\ref{knobs} shows two CDFs: one for systems that have only one known protostar within 10,000\,AU and another for systems with more than one known protostar within 10,000\,AU. The middle panel shows two CDFs based for protostars with $T_{\rm{bol}}$ above and below 50\,K, where lower $T_{\rm{bol}}$ indicates younger protostars. The right panel shows two CDFs based on the \od\ pixel value at the \citet{Tobin2016} protostellar location for \od\ above and below 0.016. Protostars at locations of higher \od\ are more likely to be in their natal star-forming filament. We select delimitations of  $T_{\rm{bol}}$~=~50\,K and \od~=~0.016 so that roughly half of the sample is in each CDF. We note that $T_{\rm{bol}}$~=~70\,K is typically used to separate Class~0 and Class~I protostars \citep{Chen1995,Enoch2009}.


Since the distribution of \gaf\ angles is separated into two CDFs for each panel in Figure~\ref{knobs}, statistically differentiating the distributions from random and perpendicular Monte Carlo simulations is more difficult. Table \ref{tab:pvals} shows that none of these CDFs can be distinguished from a random distribution, and several CDFs are statistically inconsistent with perpendicular. While we only show the corresponding $p$-value results if we use the \texttt{FILFINDER} algorithm, the results would be qualitatively the same if we used the filament fits from SExtractor. 

We also find that the empirical CDFs in each panel are not inconsistent with each other. Specifically, the $p$-value between singles and multiples is 0.80, between the two $T_{\rm{bol}}$ bins is 0.56, and between the two \od\ bins is 0.24. The latter shows that \od\ could possibly be the best discriminator between two populations of $\gamma$. This would imply that protostars that are less embedded (and likely older) have outflows perpendicular to their natal filaments. Indeed, this idea is supported by the fact that higher $T_{\rm{bol}}$ (i.e., older protostars) are closer to the perpendicular curve (albeit, very slightly) than sources with lower $T_{\rm{bol}}$. However, we stress that this trend is only tentative, as it is far from being statistically significant to draw firm conclusions. A much larger sample of protostars would allow for a better understanding of whether or not individual protostellar characteristics affect the observed $\gamma$ distribution.

\section{Discussion}\label{discussion}
We find that the observed distribution of the projected angle between outflow and filaments, $\gamma$, is significantly inconsistent with projected ``only parallel" (angles between 0$^\circ$ and 20$^\circ$) and ``only perpendicular" (angles between 70$^\circ$ and 90$^\circ$) angle distributions. The observed $\gamma$ distribution instead appears more consistent with a random distribution and for certain bimodal distributions of parallel and perpendicular angles. The best match for the bimodal distribution are angles that are only parallel 22\% of the time and only perpendicular 78\% of the time. These results are at apparent disagreement with \citet{Anathpindika2008}, but that study has a number of caveats, as explained in Appendix~\ref{anathdiscuss}. Therefore, we believe that, at least in Perseus, our results are a better representation of the actual $\gamma$ distribution.

\citet{Davis2009} also found an apparently random alignment when comparing molecular hydrogen outflows to the filament/core directions in Orion, but they did not test the idea of a mixed distribution of only parallel and only perpendicular angles. Such random alignment is supported by \citet{Tatematsu2016}, who found that the angular momentum axes of cores in the Orion~A filament are random with respect to the filamentary structure. Our study and these studies show that protostellar outflows in both low- and high-mass star-forming regions show no preferred orientation relative to their local filament. In a study that does not compare outflows angles to filaments, \citet{Ioannidis2012} investigated whether outflows are perpendicular to the Galactic plane. Specifically, they observed molecular hydrogen outflows within part of the Galactic plane ($18^\circ < l < 30^\circ$; $-1.5^\circ < b < +1.5^\circ$), and they also found a somewhat random distribution of outflow PAs, with a marginal preference for outflows to be aligned perpendicular to the Galactic plane.


Theoretical models and simulations at parsec-scales have shown that filaments can be the result of colliding clouds or flows, and the initial orientation of the angular momentum in these systems can dictate how angular momentum is transported to smaller scales. Theoretical expectations of $\gamma$ vary significantly and can often depend on the initial conditions set in the simulation. Hydrodynamic turbulent simulations of collapsing clouds by \citet{Tilley2004} show that cores within filaments can form at oblique shocks, and these shocks can impart angular momentum to the core. Simulations by \citet{Clarke2017} show that filaments accreting from a turbulent medium have a vorticity (and hence, angular momentum) that is typically parallel to filaments, which is primarily derived from radial inhomogeneous accretion.  \citet{ChenOstriker2014,ChenOstriker2015} included magnetohydrodynamics in their simulations and found that for filaments forming due to converging flows, mass flows along magnetic field lines to both the filaments and cores (which form simultaneously). For dense filaments of size-scales on order of 0.1 pc, some observations have suggested that magnetic field lines are perpendicular to the filament's elongation \citep[e.g.,][]{Matthews2000,Pereyra2004,Santos2016}. If such fields help drive gas perpendicular to the filaments, the results from \citet{Clarke2017} suggest that this could induce a vorticity parallel to the filaments. The ability for such vorticity to be transferred to angular momentum at the core scale or smaller is unclear, and this was not investigated by \citet{Clarke2017}. However, if angular momentum is inherited by the protostar in the same direction of the vorticity, we would expect the rotation of the protostar to be parallel with the filament. Indeed, simulations by \citet{Tilley2004} and \citet{Banerjee2006} show that for filaments forming due to colliding flows, oblique shocks can impart net rotation parallel to the filament, which in turn can produce parallel filaments and protostellar rotation axes. However, numerical simulations by \citet{Whitworth1995} suggest that filaments can form via two colliding clumps, and the initial net angular momentum of the system will typically be perpendicular to the filaments that form. The protostar can inherent this angular momentum, and thus its rotation axis will tend to be perpendicular to the filament. Theoretical predictions of rotation axes either parallel or perpendicular to the filament axes are at odds with observations at both the core \citep{Tatematsu2016} and protostellar scales (this study). 

Since filaments may be created through a variety of mechanisms, a combination of these mechanisms could cause outflow-filament alignment to appear more randomly aligned. Assuming the alignment is not purely random, our observations suggest that outflows are more likely to form perpendicular than parallel to the filamentary elongation. Unfortunately, two-dimensional projections of 3-dimensionally random and mostly perpendicular distributions look quite similar, making it difficult for even large samples to distinguish between the two. Moreover, the fact that the angles between outflows and filaments are neither purely parallel nor purely perpendicular may reflect how material is funneled toward the protostars at both the large and small scales.  On large scales, \citet{ChenOstriker2014} suggested that material flows along magnetic field lines, which could be mainly perpendicular to the filament along its exterior and parallel within the interior. This mix of flows could induce a more random-like vorticity to the parental cores of the protostars.

Higher resolution simulations have explored angular momentum transfer within cores (i.e., scales $\lesssim$0.1~pc). \citet{Walch2010} used smoothed particle hydrodynamic simulations of a low-mass, transonically turbulent core, and found that the rotation axes of protostars tend to be perpendicular to ``small" filaments (diameters $\sim$0.01\,pc) within cores. However, the $Herschel$-derived \od\ maps (36$\arcsec$~=~0.04\,pc resolution) do not resolve these small filaments. Observations of molecular line \citep[e.g.,]{Hacar2013} or continuum tracers \citep[e.g.,][]{Pineda2011b} suggest that filaments break into smaller substructures, and therefore the initial conditions for protostellar rotation and collapse may be set by these smaller structures. These substructures sometimes have similar elongation as their parent filaments \citep{Pineda2011b,Hacar2013}, but not always \citep[e.g.,][]{Pineda2010,Pineda2015}. At scales of $\sim$10,000~AU, elongated, flattened envelopes are observed to be perpendicular to their outflows \citep[e.g.,][]{Looney2007}. The typical size of these flattened structures and their universality remains unclear. Observational surveys that probe dense structures at scales between $\sim$0.01 to 0.1\,pc can uncover whether and at what scale an elongated structure is perpendicular with a protostar's angular momentum axis.


Regardless of the initial conditions that create filaments, the actual spin of a protostar may be independent of the filamentary structure. The local vorticity of turbulence may determine the spin of the parent core \citep{McKeeOstriker2007}. Even within the core, the rotation axes of protostars may change. \citet{Offner2016} and \citet{LeeJoyce2017} found that both turbulent accretion onto a protostar and interaction with companions can cause a significant evolution in a protostar's spin. Essentially, at small scales it is feasible that the underlying structure, turbulence, and/or multiplicity could significantly alter the initial rotation axes. While random alignment is favored in some models of turbulent accretion, even models with strong magnetic fields could result in random alignment. \citet{Mouschovias1985} suggested that for fragments linked by strong magnetic fields, the angular momentum orientation of the fragments depends solely on the shape of the magnetic flux tubes, which can have quite irregular shapes. If fragments in filaments are indeed magnetically linked, our study suggests that the flux tubes connecting them are indeed irregular.  Theoretical simulations have begun to incorporate gravity, turbulence, magnetic fields, and outflows to study the formation of filamentary complexes \citep[e.g.,][]{Myers2014,Federrath2016}. Such simulations can supply a more robust expectation of the observed distribution of $\gamma$ for a large sample of outflows and filaments.


\section{Summary}\label{summary}
The MASSES survey observed CO(2--1) in all the known Class 0/I protostars in the Perseus molecular cloud. With these data, along with ancillary observations of CO rotational transitions, we were able to determine the outflow PAs for each protostar. We compare these angles to the filament directions based on optical depth maps derived from $Herschel$ \citep{Zari2016}. We find that:
\begin{enumerate}
\item The outflow directions are randomly distributed in the Perseus molecular cloud. This random distribution appears to hold regardless of parental clump of a protostar.
\item The projected angle between the outflow and filament, $\gamma$, is significantly inconsistent with a ``purely parallel" and a ``purely perpendicular" distribution of projected angles.
\item The observed $\gamma$ distribution cannot be distinguished from a random distribution. 
\item We also consider bimodal distributions, and find a slightly more consistent distribution to the observed gamma distribution when 22\% of the projected angles are parallel and 78\% are perpendicular. Our observations are unlikely to come from bimodal distributions that are more than $\sim$33\% parallel or more than $\sim$90\% perpendicular.
\item Regardless of the multiplicity, $T_{\rm{bol}}$ (age), or opacity of the individual protostars, the observed $\gamma$ distribution cannot be distinguished from a random distribution. However, to better test how these different parameters of the protostars affect the $\gamma$ distribution, a larger sample is needed.
\end{enumerate}

We discuss the implications of the fact that outflows and filaments are neither purely perpendicular or purely parallel. We suggest that this feature could reflect the physical conditions at large or small scale. At large scale, a dominant flow direction toward cores may not exist. At small scale, the underlying structure, turbulence, and/or multiplicity could affect the angular momentum axes. Observational surveys of dust emission at scales between $\sim$0.01 to 0.1\,pc are needed to reveal whether and how a protostar's angular momentum axis may be related to its natal structure.

\acknowledgements
We thank an anonymous referee for thorough and helpful reviews.
I.W.S. acknowledges support from NASA grant NNX14AG96G. 
E.I.V. acknowledges support form the Russian Ministry of Education and Science grant 3.5602.2017.
J.J.T. acknowledges support from the University of Oklahoma, the Homer L. Dodge endowed chair, and grant 639.041.439 from the Netherlands Organisation for Scientific Research (NWO).  
J.E.P. acknowledges the financial support of the European Research Council (ERC; project PALs 320620).
The authors thank the SMA staff for executing these observations as part of the queue schedule, Charlie Qi and Mark Gurwell for their technical assistance with the SMA data, and Eric Keto for his guidance with SMA large-scale projects.
The Submillimeter Array is a joint project between the Smithsonian Astrophysical Observatory and the Academia Sinica Institute of Astronomy and Astrophysics and is funded by the Smithsonian Institution and the Academia Sinica.
This research has made use of the VizieR catalogue access tool and the SIMBAD database operated at CDS, Strasbourg, France. 
This research made use of APLpy, an open-source plotting package for Python \citep{Robitaille2012}.

\appendix
\section{A. Monte Carlo simulations}\label{montecarlo}
Many studies have used Monte Carlo simulations to show the expected observed distribution of angles of two vectors projected into three dimensions. Several of these studies \citep{Hull2013,Hull2014,Lee2016,Offner2016} were specifically interested in the same projected distributions we are interested in this study, i.e., the projection of angles that are 3-dimensionally purely parallel (between 0 and 20$^\circ$), purely perpendicular (70--90$^\circ$), or completely random (0--90$^\circ$). These studies do not discuss the exact details of the Monte Carlo simulations. Here we discuss our Monte Carlo method, and the results are consistent with the aforementioned studies.

For our methodology, we generated $N$ pairs of 3-dimensional vectors with each vector random about the sky. To generate a random vector, we chose a random point on the surface of a unit sphere and then connected the sphere's origin to this point. For the purposes of Monte Carlo simulations, sampling a random point from a unit sphere that avoids biases has been well-studied \citep[e.g.,][]{Marsaglia1972}. We outline one such way to select random points on a unit sphere below, which is based on \citet{Weisstein2017}. We first selected a random angle $\theta$ between 0 and 2$\pi$ and a random number $u$ that's between --1 and 1. From random variables $\theta$ and $u$, we then selected a random point on a unit sphere at position $x$, $y$, and $z$ where

\begin{equation}
x = \sqrt{1-u^2}~\rm{cos}\,\theta
\end{equation}
\begin{equation}
y = \sqrt{1-u^2}~\rm{sin}\,\theta  \\
\end{equation}
\begin{equation}
z = u.
\end{equation}


A unit vector between the sphere's origin and this point is:
\begin{equation}
\vec{v} = \left[ \begin{array}{c} x \\ y \\ z \end{array} \right]
\end{equation}

\begin{figure}[ht!]
\begin{center}
\includegraphics[width=1\columnwidth]{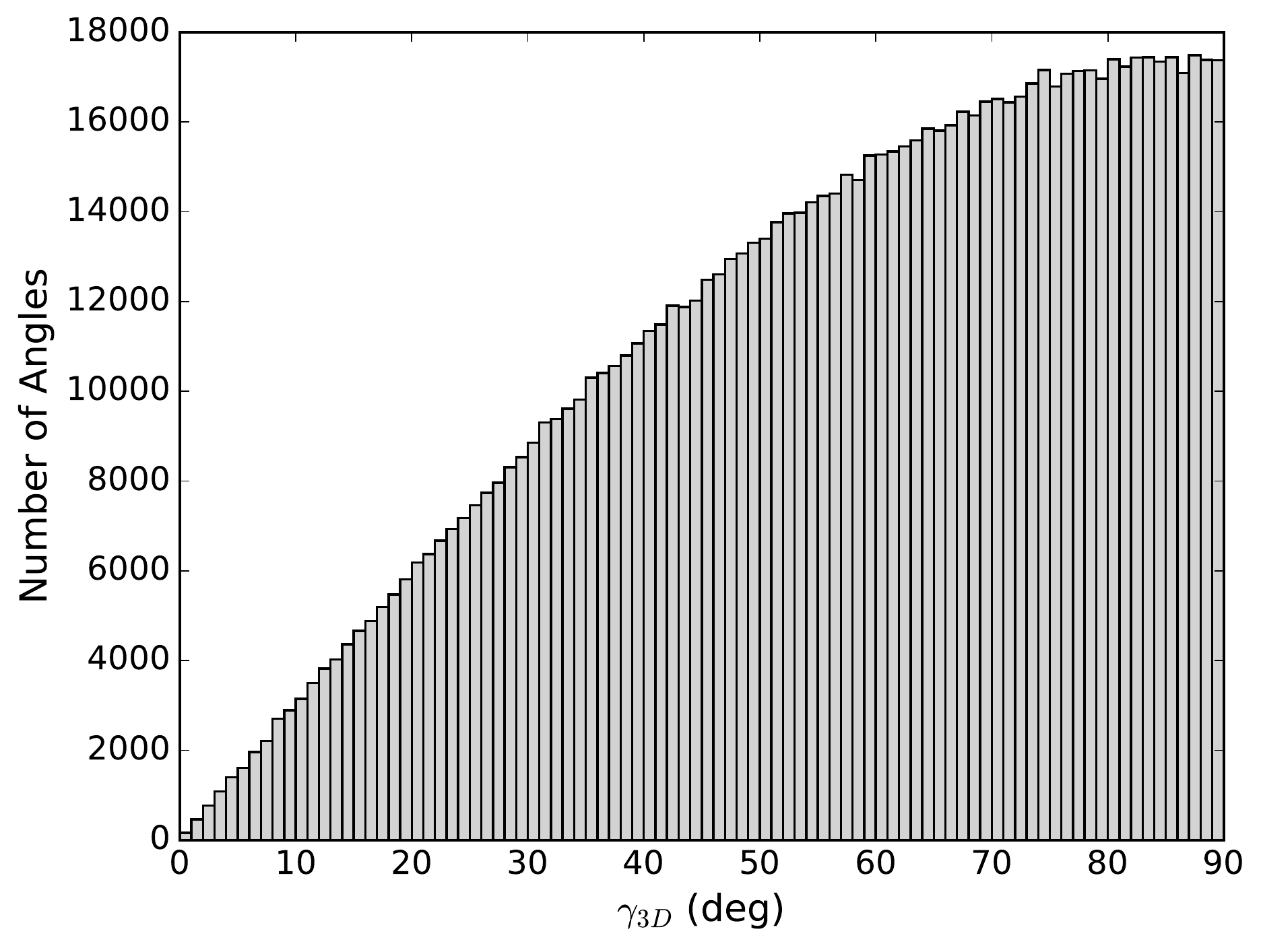}
\end{center}
\caption{Histogram of $\gamma_{3\rm{D}}$ for a Monte Carlo simulation of $N = 10^6$ vector pairs. Histogram bin widths are 1$^\circ$. This histogram shows the approximate shape of the distribution of all possible angles between two vectors in a unit sphere.
}
\label{gamma3ddistribution} 
\end{figure}

To randomly sample from all angles within a unit sphere, we generated two random unit vectors, $\vec{v}_1$ and $\vec{v}_2$ and measure the angle between the vectors. The angle is simply 

\begin{equation}
\gamma_{3\rm{D}} = \mbox{arccos}(\vec{v_1} \cdot \vec{v_2}).
\end{equation}

Since we are interested in the smallest angle created by the two intersecting vectors, we constrained $\gamma_{3\rm{D}}$  to be between 0$^\circ$ and 90$^\circ$, e.g., if $\gamma_{3\rm{D}}$ is larger than 90$^\circ$, we subtracted $\gamma_{3\rm{D}}$ from 180$^\circ$. We generated $N$ pairs of vectors to produce $N$ angles of $\gamma_{3\rm{D}}$. For the Monte Carlo simulations in this paper, we chose $N=10^6$. We show the distribution of $\gamma_{3\rm{D}}$ for $N=10^6$ via the histogram in Figure~\ref{gamma3ddistribution}. We then mapped each $\gamma_{3\rm{D}}$ angle to a projected angle in 2D, $\gamma$, by setting one axis for the vector pair to 0 (the $x$-value of the vector in our code) and calculating the new angle between the vectors.

From this mapping, we can extract a range of angles from the distribution of $\gamma_{3\rm{D}}$ and plot its corresponding $\gamma$ distribution. For this study, we were primarily interested in projections for 3-dimensional angles that are purely parallel (between 0 and 20$^\circ$), purely perpendicular (70--90$^\circ$), or completely random (0--90$^\circ$). For the Monte Carlo sample size of $N=10^6$ (equivalent to the number for the completely random sample size), we extracted from the $\gamma_{3\rm{D}}$ distribution $\sim$60,000 projections for a purely parallel sample and $\sim$340,000 for a purely perpendicular sample. The reason why the sample size for purely perpendicular is much larger than purely parallel is simply due to the fact that perpendicular-like angles are much more likely for two random vectors in a unit sphere (Figure \ref{gamma3ddistribution}). Our tests show that the curve of the CDF of the Monte Carlo simulation (e.g., Figure~\ref{main_cdf}) is very smooth as long as the sample size is larger than $\sim$20,000 projections.


\section{B. Discrepancy with Anathpindika \& Whitworth}\label{anathdiscuss}
As seen in Table \ref{tab:pvals} and Figure~\ref{sextractor}, \citet[][henceforth in this appendix, \citetalias{Anathpindika2008}]{Anathpindika2008} found a distribution of projected outflow-filament angles, $\gamma$, that favors outflows and filaments that are generally perpendicular rather than random. When comparing a random distribution to the \citetalias{Anathpindika2008} distribution of $\gamma$, the AD test $p$-value is 0.017, indicating a significantly non-random distribution. \citetalias{Anathpindika2008} also found that, if they assumed $\gamma$ follows a tapered Gaussian (i.e., between 0$^\circ$ and 90$^\circ$) centered at perpendicular, 72\% of the time the outflow is within 45$^\circ$ of being perpendicular to the filament. 


To identify the PA of the outflow, \citetalias{Anathpindika2008} connected a line between a near-IR identified YSO and the corresponding Herbig Haro Object from \citet{Reipurth1999}. The PA of the filaments are determined from flux maps of various submillimeter surveys using SExtractor in STARLINK (with a visual confirmation of the PA). \citetalias{Anathpindika2008} acknowledged a few selection effects that may bias their results. Specifically, they assumed that all objects have random inclinations, although adjacent sources may have correlated inclinations. Our study also suffers from this bias. \citetalias{Anathpindika2008} also suggested that they are inherently more likely to find perpendicular outflows since Herbig Haro objects are more likely to be extincted if they are coincident with the filament. For these reasons, they call their conclusion not statistically robust.

\begin{figure}[ht!]
\begin{center}
\includegraphics[width=1\columnwidth]{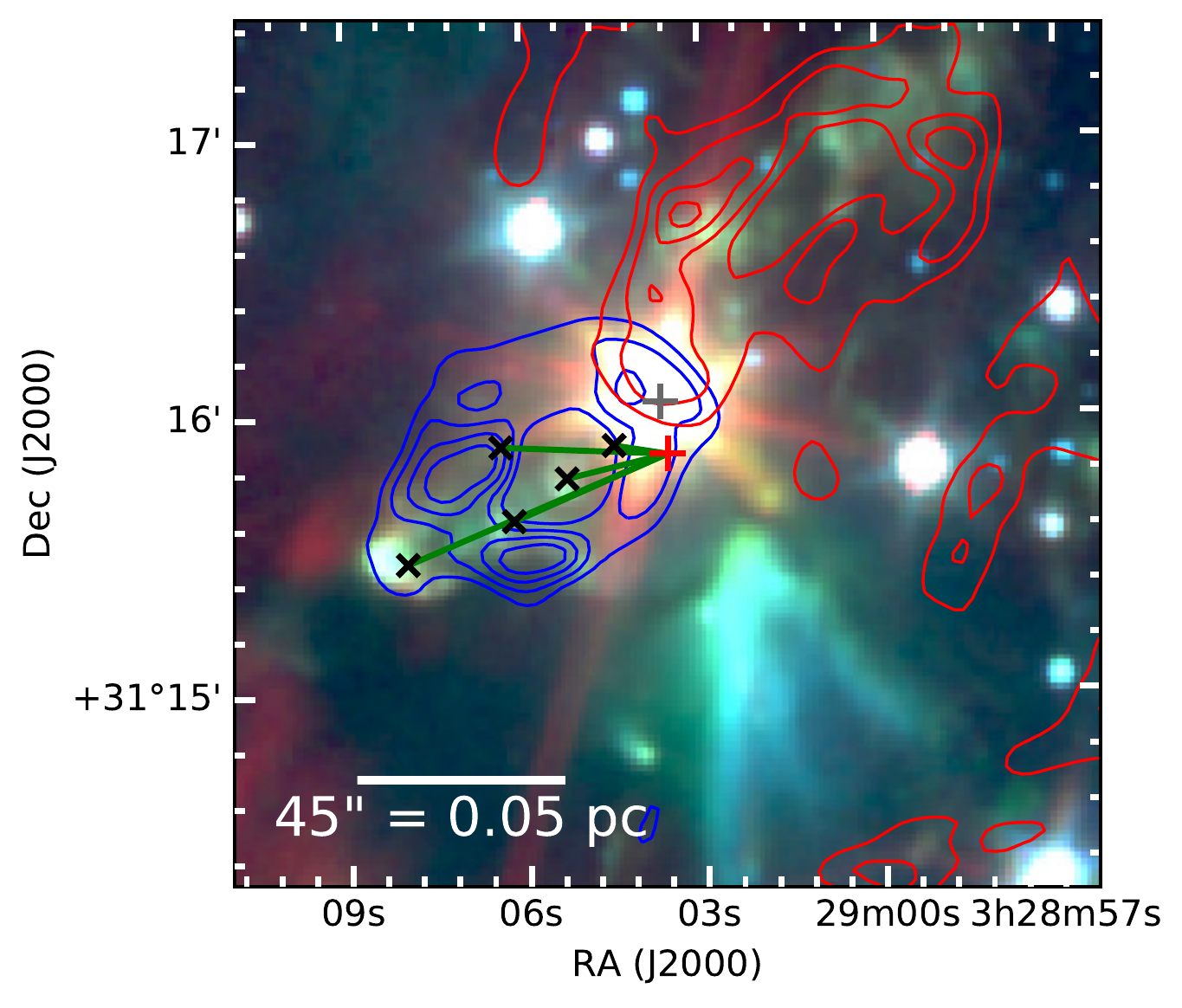}
\end{center}
\caption{$Spitzer$ three-color image of the SVS~13 star-forming region in NGC~1333, with blue, green, and red showing 3.6, 4.5, and 8.0\,$\mu$m, respectively. The CO(1--0) integrated intensity maps from \citet{Plunkett2013} are shown as blue and red contours for the blue and red lobes, respectively. The red plus sign shows the \citet{Anathpindika2008} position of the protostar based on published near-IR data, while the gray plus sign shows the actual position of the outflow ejection from SVS~13A \citep[based on 3mm continuum emission,][]{Plunkett2013}. Black crosses show the locations of Herbig Haro objects identified in \citet{Reipurth1999}. The green lines show 5 outflows identified in \citet{Anathpindika2008}, although probably only one outflow produces all these Herbig Haro objects \citep{Plunkett2013}.
}
\label{anath} 
\end{figure}

\citetalias{Anathpindika2008} also have some other disadvantages with their dataset. Their measured outflow angles rely primarily on published catalogs rather than the physical images. For about half of their sources, they interpreted multiple Herbig Haro objects emitting from a young stellar object as independent outflows. However, upon further analysis, we find this interpretation is not always accurate. As an example, Figure~\ref{anath} shows a 3-color $Spitzer$ image of the outflow emanating from the SVS~13 protostellar region. The $Spitzer$ image shows only one obvious bipolar outflow from the protostar (greenish 4.5\,$\mu$m color), and the molecular CO(1--0) line observations confirm this is a single outflow \citep[contours in Figure~\ref{anath};][]{Plunkett2013}. However, \citetalias{Anathpindika2008} declared that the five HH objects associated with this outflow are five separate outflows, and each of these had a measurement of $\gamma$ above 45$^\circ$. Therefore, \citetalias{Anathpindika2008} sometimes have multiple measurements for $\gamma$ for a single outflow, which will significantly bias their results toward a non-random distribution. Moreover, Figure~\ref{anath} shows that significantly different measurements for \pao\ can be made for each Herbig Haro object for the same outflow. The dispersion of HH objects about the outflow lobe may occur due to a precessing outflow coupled with episodic ejections \citep[e.g.,][]{ArceGoodman2001,Arce2010} and/or due to the structure (e.g., clumpiness) of the ambient cloud. Therefore, measuring \pao\ from Herbig Haro objects alone can result in large \pao\ measurement errors. \citetalias{Anathpindika2008} also rely on previous published protostellar positions based on near-IR observations, and these objects sometimes are not the source of the outflows. Figure~\ref{anath} shows an example of this protostellar misidentification, where \citetalias{Anathpindika2008} use the location of SVS~13B as the origin of the outflow (marked in red), whereas interferometry data from \citet{Plunkett2013} show that the outflow originates from SVS~13A (marked in blue; coincides with 3mm continuum emission). \citetalias{Anathpindika2008}'s errant location of the protostar causes the outflow PA to be mis-measured in SVS~13 by up to $\sim$50$^\circ$.

Given that \citetalias{Anathpindika2008} have these significant biases and shortcomings in their study, we believe the tentative evidence by \citetalias{Anathpindika2008} of preferentially perpendicular (and non-random) outflows and filaments is unreliable.



\bibliography{stephens_bib}

\begin{thebibliography}{}
\expandafter\ifx\csname natexlab\endcsname\relax\def\natexlab#1{#1}\fi

\bibitem[{{Abreu-Vicente} {et~al.}(2016){Abreu-Vicente}, {Stutz}, {Robitaille},
  {Henning}, \& {Keto}}]{Abreu-Vicente2016}
{Abreu-Vicente}, J., {Stutz}, A., {Robitaille}, T., {Henning}, T., \& {Keto},
  E. 2016, ArXiv e-prints, arXiv:1605.03195

\bibitem[{{Anathpindika} \& {Whitworth}(2008)}]{Anathpindika2008}
{Anathpindika}, S., \& {Whitworth}, A.~P. 2008, \aap, 487, 605

\bibitem[{{Andr{\'e}} {et~al.}(2010){Andr{\'e}}, {Men'shchikov}, {Bontemps},
  {K{\"o}nyves}, {Motte}, {Schneider}, {Didelon}, {Minier}, {Saraceno},
  {Ward-Thompson}, {di Francesco}, {White}, {Molinari}, {Testi}, {Abergel},
  {Griffin}, {Henning}, {Royer}, {Mer{\'{\i}}n}, {Vavrek}, {Attard},
  {Arzoumanian}, {Wilson}, {Ade}, {Aussel}, {Baluteau}, {Benedettini},
  {Bernard}, {Blommaert}, {Cambr{\'e}sy}, {Cox}, {di Giorgio}, {Hargrave},
  {Hennemann}, {Huang}, {Kirk}, {Krause}, {Launhardt}, {Leeks}, {Le Pennec},
  {Li}, {Martin}, {Maury}, {Olofsson}, {Omont}, {Peretto}, {Pezzuto}, {Prusti},
  {Roussel}, {Russeil}, {Sauvage}, {Sibthorpe}, {Sicilia-Aguilar}, {Spinoglio},
  {Waelkens}, {Woodcraft}, \& {Zavagno}}]{Andre2010}
{Andr{\'e}}, P., {Men'shchikov}, A., {Bontemps}, S., {et~al.} 2010, \aap, 518,
  L102

\bibitem[{{Arce} {et~al.}(2010){Arce}, {Borkin}, {Goodman}, {Pineda}, \&
  {Halle}}]{Arce2010}
{Arce}, H.~G., {Borkin}, M.~A., {Goodman}, A.~A., {Pineda}, J.~E., \& {Halle},
  M.~W. 2010, \apj, 715, 1170

\bibitem[{{Arce} \& {Goodman}(2001)}]{ArceGoodman2001}
{Arce}, H.~G., \& {Goodman}, A.~A. 2001, \apj, 554, 132

\bibitem[{{Arzoumanian} {et~al.}(2011){Arzoumanian}, {Andr{\'e}}, {Didelon},
  {K{\"o}nyves}, {Schneider}, {Men'shchikov}, {Sousbie}, {Zavagno}, {Bontemps},
  {di Francesco}, {Griffin}, {Hennemann}, {Hill}, {Kirk}, {Martin}, {Minier},
  {Molinari}, {Motte}, {Peretto}, {Pezzuto}, {Spinoglio}, {Ward-Thompson},
  {White}, \& {Wilson}}]{Arzoumanian2011}
{Arzoumanian}, D., {Andr{\'e}}, P., {Didelon}, P., {et~al.} 2011, \aap, 529, L6

\bibitem[{{Auddy} {et~al.}(2016){Auddy}, {Basu}, \& {Kudoh}}]{Auddy2016}
{Auddy}, S., {Basu}, S., \& {Kudoh}, T. 2016, \apj, 831, 46

\bibitem[{{Banerjee} {et~al.}(2006){Banerjee}, {Pudritz}, \&
  {Anderson}}]{Banerjee2006}
{Banerjee}, R., {Pudritz}, R.~E., \& {Anderson}, D.~W. 2006, \mnras, 373, 1091

\bibitem[{{Bertin} \& {Arnouts}(1996)}]{Bertin1996}
{Bertin}, E., \& {Arnouts}, S. 1996, \aaps, 117, 393

\bibitem[{Blum(1967)}]{Blum1967}
Blum, H. 1967, in Models for the Perception of Speech and Visual Form, ed.
  W.~Wathen-Dunn (Cambridge: MIT Press), 362--380

\bibitem[{{Bodenheimer}(1995)}]{Bodenheimer1995}
{Bodenheimer}, P. 1995, \araa, 33, 199

\bibitem[{{Chen} \& {Ostriker}(2014)}]{ChenOstriker2014}
{Chen}, C.-Y., \& {Ostriker}, E.~C. 2014, \apj, 785, 69

\bibitem[{{Chen} \& {Ostriker}(2015)}]{ChenOstriker2015}
---. 2015, \apj, 810, 126

\bibitem[{{Chen} {et~al.}(1995){Chen}, {Myers}, {Ladd}, \& {Wood}}]{Chen1995}
{Chen}, H., {Myers}, P.~C., {Ladd}, E.~F., \& {Wood}, D.~O.~S. 1995, \apj, 445,
  377

\bibitem[{{Chen} {et~al.}(2010){Chen}, {Arce}, {Zhang}, {Bourke}, {Launhardt},
  {Schmalzl}, \& {Henning}}]{Chen2010}
{Chen}, X., {Arce}, H.~G., {Zhang}, Q., {et~al.} 2010, \apj, 715, 1344

\bibitem[{{Clark} {et~al.}(2014){Clark}, {Peek}, \& {Putman}}]{Clark2014}
{Clark}, S.~E., {Peek}, J.~E.~G., \& {Putman}, M.~E. 2014, \apj, 789, 82

\bibitem[{{Clarke} {et~al.}(2017){Clarke}, {Whitworth}, {Duarte-Cabral}, \&
  {Hubber}}]{Clarke2017}
{Clarke}, S.~D., {Whitworth}, A.~P., {Duarte-Cabral}, A., \& {Hubber}, D.~A.
  2017, \mnras, 468, 2489

\bibitem[{{Davis} {et~al.}(2009){Davis}, {Froebrich}, {Stanke}, {Megeath},
  {Kumar}, {Adamson}, {Eisl{\"o}ffel}, {Gredel}, {Khanzadyan}, {Lucas},
  {Smith}, \& {Varricatt}}]{Davis2009}
{Davis}, C.~J., {Froebrich}, D., {Stanke}, T., {et~al.} 2009, \aap, 496, 153

\bibitem[{{Engmann} \& {Cousineau}(2011)}]{Engmann2011}
{Engmann}, S., \& {Cousineau}, D. 2011, Journal of Applied Quantitative
  Methods, 6(3), 1

\bibitem[{{Enoch} {et~al.}(2009){Enoch}, {Evans}, {Sargent}, \&
  {Glenn}}]{Enoch2009}
{Enoch}, M.~L., {Evans}, II, N.~J., {Sargent}, A.~I., \& {Glenn}, J. 2009,
  \apj, 692, 973

\bibitem[{{Federrath}(2016)}]{Federrath2016}
{Federrath}, C. 2016, \mnras, 457, 375

\bibitem[{{Feigelson} \& {Babu}(1992)}]{Feigelson1992}
{Feigelson}, E.~D., \& {Babu}, G.~J. 1992, \apj, 397, 55

\bibitem[{{Fogerty} {et~al.}(2016){Fogerty}, {Carroll-Nellenback}, {Frank},
  {Heitsch}, \& {Pon}}]{Fogerty2017}
{Fogerty}, E.~L., {Carroll-Nellenback}, J., {Frank}, A., {Heitsch}, F., \&
  {Pon}, A. 2016, ArXiv e-prints, arXiv:1609.02918

\bibitem[{{Goodman} {et~al.}(1993){Goodman}, {Benson}, {Fuller}, \&
  {Myers}}]{Goodman1993}
{Goodman}, A.~A., {Benson}, P.~J., {Fuller}, G.~A., \& {Myers}, P.~C. 1993,
  \apj, 406, 528

\bibitem[{{Hacar} {et~al.}(2013){Hacar}, {Tafalla}, {Kauffmann}, \&
  {Kov{\'a}cs}}]{Hacar2013}
{Hacar}, A., {Tafalla}, M., {Kauffmann}, J., \& {Kov{\'a}cs}, A. 2013, \aap,
  554, A55

\bibitem[{{Heyer}(1988)}]{Heyer1988}
{Heyer}, M.~H. 1988, \apj, 324, 311

\bibitem[{{Hirota} {et~al.}(2008){Hirota}, {Bushimata}, {Choi}, {Honma},
  {Imai}, {Iwadate}, {Jike}, {Kameya}, {Kamohara}, {Kan-Ya}, {Kawaguchi},
  {Kijima}, {Kobayashi}, {Kuji}, {Kurayama}, {Manabe}, {Miyaji}, {Nagayama},
  {Nakagawa}, {Oh}, {Omodaka}, {Oyama}, {Sakai}, {Sasao}, {Sato}, {Shibata},
  {Tamura}, \& {Yamashita}}]{Hirota2008}
{Hirota}, T., {Bushimata}, T., {Choi}, Y.~K., {et~al.} 2008, \pasj, 60, 37

\bibitem[{{Ho} {et~al.}(2004){Ho}, {Moran}, \& {Lo}}]{Ho2004}
{Ho}, P.~T.~P., {Moran}, J.~M., \& {Lo}, K.~Y. 2004, \apjl, 616, L1

\bibitem[{{Hou} {et~al.}(2009){Hou}, {Parker}, {Harris}, \& {Wilman}}]{Hou2009}
{Hou}, A., {Parker}, L.~C., {Harris}, W.~E., \& {Wilman}, D.~J. 2009, \apj,
  702, 1199

\bibitem[{{Hull} {et~al.}(2013){Hull}, {Plambeck}, {Bolatto}, {Bower},
  {Carpenter}, {Crutcher}, {Fiege}, {Franzmann}, {Hakobian}, {Heiles}, {Houde},
  {Hughes}, {Jameson}, {Kwon}, {Lamb}, {Looney}, {Matthews}, {Mundy}, {Pillai},
  {Pound}, {Stephens}, {Tobin}, {Vaillancourt}, {Volgenau}, \&
  {Wright}}]{Hull2013}
{Hull}, C.~L.~H., {Plambeck}, R.~L., {Bolatto}, A.~D., {et~al.} 2013, \apj,
  768, 159

\bibitem[{{Hull} {et~al.}(2014){Hull}, {Plambeck}, {Kwon}, {Bower},
  {Carpenter}, {Crutcher}, {Fiege}, {Franzmann}, {Hakobian}, {Heiles}, {Houde},
  {Hughes}, {Lamb}, {Looney}, {Marrone}, {Matthews}, {Pillai}, {Pound},
  {Rahman}, {Sandell}, {Stephens}, {Tobin}, {Vaillancourt}, {Volgenau}, \&
  {Wright}}]{Hull2014}
{Hull}, C.~L.~H., {Plambeck}, R.~L., {Kwon}, W., {et~al.} 2014, \apjs, 213, 13

\bibitem[{{Ioannidis} \& {Froebrich}(2012)}]{Ioannidis2012}
{Ioannidis}, G., \& {Froebrich}, D. 2012, \mnras, 421, 3257

\bibitem[{{Isobe} {et~al.}(1990){Isobe}, {Feigelson}, {Akritas}, \&
  {Babu}}]{Isobe1990}
{Isobe}, T., {Feigelson}, E.~D., {Akritas}, M.~G., \& {Babu}, G.~J. 1990, \apj,
  364, 104

\bibitem[{{Koch} \& {Rosolowsky}(2015)}]{Koch2015}
{Koch}, E.~W., \& {Rosolowsky}, E.~W. 2015, \mnras, 452, 3435

\bibitem[{{Lee} {et~al.}(2017){Lee}, {Hull}, \& {Offner}}]{LeeJoyce2017}
{Lee}, J.~W.~Y., {Hull}, C.~L.~H., \& {Offner}, S.~S.~R. 2017, \apj, 834, 201

\bibitem[{{Lee} {et~al.}(2015){Lee}, {Dunham}, {Myers}, {Tobin}, {Kristensen},
  {Pineda}, {Vorobyov}, {Offner}, {Arce}, {Li}, {Bourke}, {J{\o}rgensen},
  {Goodman}, {Sadavoy}, {Chandler}, {Harris}, {Kratter}, {Looney}, {Melis},
  {Perez}, \& {Segura-Cox}}]{Lee2015}
{Lee}, K.~I., {Dunham}, M.~M., {Myers}, P.~C., {et~al.} 2015, \apj, 814, 114

\bibitem[{{Lee} {et~al.}(2016){Lee}, {Dunham}, {Myers}, {Arce}, {Bourke},
  {Goodman}, {J{\o}rgensen}, {Kristensen}, {Offner}, {Pineda}, {Tobin}, \&
  {Vorobyov}}]{Lee2016}
---. 2016, \apjl, 820, L2

\bibitem[{{Looney} {et~al.}(2007){Looney}, {Tobin}, \& {Kwon}}]{Looney2007}
{Looney}, L.~W., {Tobin}, J.~J., \& {Kwon}, W. 2007, \apjl, 670, L131

\bibitem[{Marsaglia(1972)}]{Marsaglia1972}
Marsaglia, G. 1972, Ann. Math. Statist., 43, 645

\bibitem[{{Matthews} \& {Wilson}(2000)}]{Matthews2000}
{Matthews}, B.~C., \& {Wilson}, C.~D. 2000, \apj, 531, 868

\bibitem[{{McKee} \& {Ostriker}(2007)}]{McKeeOstriker2007}
{McKee}, C.~F., \& {Ostriker}, E.~C. 2007, \araa, 45, 565

\bibitem[{{Mouschovias} \& {Morton}(1985)}]{Mouschovias1985}
{Mouschovias}, T.~C., \& {Morton}, S.~A. 1985, \apj, 298, 205

\bibitem[{{Myers} {et~al.}(2014){Myers}, {Klein}, {Krumholz}, \&
  {McKee}}]{Myers2014}
{Myers}, A.~T., {Klein}, R.~I., {Krumholz}, M.~R., \& {McKee}, C.~F. 2014,
  \mnras, 439, 3420

\bibitem[{{Myers} {et~al.}(1991){Myers}, {Fuller}, {Goodman}, \&
  {Benson}}]{Myers1991}
{Myers}, P.~C., {Fuller}, G.~A., {Goodman}, A.~A., \& {Benson}, P.~J. 1991,
  \apj, 376, 561

\bibitem[{{Offner} {et~al.}(2016){Offner}, {Dunham}, {Lee}, {Arce}, \&
  {Fielding}}]{Offner2016}
{Offner}, S.~S.~R., {Dunham}, M.~M., {Lee}, K.~I., {Arce}, H.~G., \&
  {Fielding}, D.~B. 2016, \apjl, 827, L11

\bibitem[{{Pereyra} \& {Magalh{\~a}es}(2004)}]{Pereyra2004}
{Pereyra}, A., \& {Magalh{\~a}es}, A.~M. 2004, \apj, 603, 584

\bibitem[{{Pineda} {et~al.}(2010){Pineda}, {Goodman}, {Arce}, {Caselli},
  {Foster}, {Myers}, \& {Rosolowsky}}]{Pineda2010}
{Pineda}, J.~E., {Goodman}, A.~A., {Arce}, H.~G., {et~al.} 2010, \apjl, 712,
  L116

\bibitem[{{Pineda} {et~al.}(2011{\natexlab{a}}){Pineda}, {Goodman}, {Arce},
  {Caselli}, {Longmore}, \& {Corder}}]{Pineda2011b}
---. 2011{\natexlab{a}}, \apjl, 739, L2

\bibitem[{{Pineda} {et~al.}(2011{\natexlab{b}}){Pineda}, {Arce}, {Schnee},
  {Goodman}, {Bourke}, {Foster}, {Robitaille}, {Tanner}, {Kauffmann},
  {Tafalla}, {Caselli}, \& {Anglada}}]{Pineda2011}
{Pineda}, J.~E., {Arce}, H.~G., {Schnee}, S., {et~al.} 2011{\natexlab{b}},
  \apj, 743, 201

\bibitem[{{Pineda} {et~al.}(2015){Pineda}, {Offner}, {Parker}, {Arce},
  {Goodman}, {Caselli}, {Fuller}, {Bourke}, \& {Corder}}]{Pineda2015}
{Pineda}, J.~E., {Offner}, S.~S.~R., {Parker}, R.~J., {et~al.} 2015, \nat, 518,
  213

\bibitem[{{Plunkett} {et~al.}(2013){Plunkett}, {Arce}, {Corder}, {Mardones},
  {Sargent}, \& {Schnee}}]{Plunkett2013}
{Plunkett}, A.~L., {Arce}, H.~G., {Corder}, S.~A., {et~al.} 2013, \apj, 774, 22

\bibitem[{{Razali} \& {Wah}(2011)}]{Razali2011}
{Razali}, N.~M., \& {Wah}, Y.~B. 2011, Journal of Statistical Modeling and
  Analytics, 2(1), 21

\bibitem[{{Reipurth}(1999)}]{Reipurth1999}
{Reipurth}, B. 1999, A general catalogue of Herbig-Haro objects, 2. edition,
  \url{http://casa.colorado.edu/hhcat}

\bibitem[{{Robitaille} \& {Bressert}(2012)}]{Robitaille2012}
{Robitaille}, T., \& {Bressert}, E. 2012, {APLpy: Astronomical Plotting Library
  in Python}, Astrophysics Source Code Library, ascl:1208.017

\bibitem[{{Sadavoy} {et~al.}(2014){Sadavoy}, {Di Francesco}, {Andr{\'e}},
  {Pezzuto}, {Bernard}, {Maury}, {Men'shchikov}, {Motte}, {Nguyen-Lu'o'ng},
  {Schneider}, {Arzoumanian}, {Benedettini}, {Bontemps}, {Elia}, {Hennemann},
  {Hill}, {K{\"o}nyves}, {Louvet}, {Peretto}, {Roy}, \& {White}}]{Sadavoy2014}
{Sadavoy}, S.~I., {Di Francesco}, J., {Andr{\'e}}, P., {et~al.} 2014, \apjl,
  787, L18

\bibitem[{{Santos} {et~al.}(2016){Santos}, {Busquet}, {Franco}, {Girart}, \&
  {Zhang}}]{Santos2016}
{Santos}, F.~P., {Busquet}, G., {Franco}, G.~A.~P., {Girart}, J.~M., \&
  {Zhang}, Q. 2016, \apj, 832, 186

\bibitem[{{Sault} {et~al.}(1995){Sault}, {Teuben}, \& {Wright}}]{Sault1995}
{Sault}, R.~J., {Teuben}, P.~J., \& {Wright}, M.~C.~H. 1995, in Astronomical
  Society of the Pacific Conference Series, Vol.~77, Astronomical Data Analysis
  Software and Systems IV, ed. R.~A. {Shaw}, H.~E. {Payne}, \& J.~J.~E.
  {Hayes}, 433

\bibitem[{Stephens(1974)}]{Stephens1974}
Stephens, M.~A. 1974, Journal of the American Statistical Association, 69, 730

\bibitem[{{Tatematsu} {et~al.}(2016){Tatematsu}, {Ohashi}, {Sanhueza}, {Nguyen
  Luong}, {Umemoto}, \& {Mizuno}}]{Tatematsu2016}
{Tatematsu}, K., {Ohashi}, S., {Sanhueza}, P., {et~al.} 2016, \pasj, 68, 24

\bibitem[{{Tilley} \& {Pudritz}(2004)}]{Tilley2004}
{Tilley}, D.~A., \& {Pudritz}, R.~E. 2004, \mnras, 353, 769

\bibitem[{{Tobin} {et~al.}(2016){Tobin}, {Looney}, {Li}, {Chandler}, {Dunham},
  {Segura-Cox}, {Sadavoy}, {Melis}, {Harris}, {Kratter}, \&
  {Perez}}]{Tobin2016}
{Tobin}, J.~J., {Looney}, L.~W., {Li}, Z.-Y., {et~al.} 2016, \apj, 818, 73

\bibitem[{{Walch} {et~al.}(2010){Walch}, {Naab}, {Whitworth}, {Burkert}, \&
  {Gritschneder}}]{Walch2010}
{Walch}, S., {Naab}, T., {Whitworth}, A., {Burkert}, A., \& {Gritschneder}, M.
  2010, \mnras, 402, 2253

\bibitem[{{Weisstein}(2017)}]{Weisstein2017}
{Weisstein}, E.~W. 2017, {From MathWorld--A Wolfram Web Resource.
  \url{http://mathworld.wolfram.com/SpherePointPicking.html}}

\bibitem[{{Whitworth} {et~al.}(1995){Whitworth}, {Chapman}, {Bhattal},
  {Disney}, {Pongracic}, \& {Turner}}]{Whitworth1995}
{Whitworth}, A.~P., {Chapman}, S.~J., {Bhattal}, A.~S., {et~al.} 1995, \mnras,
  277, 727

\bibitem[{{Young} {et~al.}(2015){Young}, {Young}, {Lai}, {Dunham}, \&
  {Evans}}]{Young2015}
{Young}, K.~E., {Young}, C.~H., {Lai}, S.-P., {Dunham}, M.~M., \& {Evans}, II,
  N.~J. 2015, \aj, 150, 40

\bibitem[{{Zari} {et~al.}(2016){Zari}, {Lombardi}, {Alves}, {Lada}, \&
  {Bouy}}]{Zari2016}
{Zari}, E., {Lombardi}, M., {Alves}, J., {Lada}, C.~J., \& {Bouy}, H. 2016,
  \aap, 587, A106

\end{thebibliography}

\section{}
\newpage

\clearpage
\begin{turnpage}
\begin{deluxetable*}{lccl@{}ccccccccccccccccc}
\tabletypesize{\tiny}
\tablecaption{Source Information and Measured Position Angles \label{tab:angles}}
\tablehead{\colhead{\tiny Source} & \colhead{\tiny RA\tablenotemark{b}} & \colhead{\tiny DEC\tablenotemark{b}} & \colhead{\tiny Other Names\tablenotemark{c}} & \colhead{\tiny Multiple} & \colhead{\tiny $T_{\rm{bol}}$} & \colhead{\tiny \od} & \colhead{\tiny Blue PA} & \colhead{\tiny Red PA} & \colhead{\tiny \pao} & \colhead{\tiny Ref/Info} & \colhead{\tiny \pafilf} & \colhead{\tiny \gaf}  & \colhead{\tiny \pafone} & \colhead{\tiny \paftwo} & \colhead{\tiny \pafthree} & \colhead{\tiny \paffour} & \colhead{\tiny \paffive} & \colhead{\tiny \pafsix} & \colhead{\tiny $\gamma_{\mbox{se,S}}$} & \colhead{\tiny $\gamma_{\mbox{se,L}}$}\\
\colhead{\tiny Name\tablenotemark{a}} & \colhead{\tiny (J2000)} & \colhead{\tiny (J2000)} & & \colhead{\tiny (Y/N)} & \colhead{\tiny (K)} & \tiny ($\times 10^3$) & \colhead{\tiny ($^\circ$)} & \colhead{\tiny ($^\circ$)} & \colhead{\tiny ($^\circ$)} & & 
\colhead{\tiny ($^\circ$)} & \colhead{\tiny ($^\circ$)} & \colhead{\tiny ($^\circ$)} & \colhead{\tiny ($^\circ$)} & \colhead{\tiny ($^\circ$)} & \colhead{\tiny ($^\circ$)} & \colhead{\tiny ($^\circ$)} & \colhead{\tiny ($^\circ$)} & \colhead{\tiny ($^\circ$)} & \colhead{\tiny ($^\circ$)}
}
\startdata
\tiny Per-emb-1 & \tiny 03:43:56.806 & \tiny +32:00:50.202 & \tiny HH211-MMS & \tiny N & \tiny 27 & \tiny 2.2 & \tiny 114 & \tiny -61 & \tiny 116 & \tiny (1) & \tiny 40 & \tiny 76 & \tiny 82 & \tiny 35 & \tiny 25 & \tiny 24 & \tiny 19 & \tiny 21 & \tiny 79 & \tiny 39 \\
\tiny Per-emb-2 & \tiny 03:32:17.932 & \tiny +30:49:47.705 & \tiny IRAS~03292+3039 & \tiny Y & \tiny 27 & \tiny 2.4 & \tiny 129 & \tiny -50 & \tiny 129 & \tiny (1) & \tiny 132 & \tiny 3 & \tiny 4 & \tiny 77 & \tiny 83 & \tiny 85 & \tiny 87 & \tiny 86 & \tiny 75 & \tiny 85 \\
\tiny Per-emb-3 & \tiny 03:29:00.575 & \tiny +31:12:00.204 & \tiny ... & \tiny N & \tiny 32 & \tiny 1.1 & \tiny -82 & \tiny 95 & \tiny 97 & \tiny (2) & \tiny 10 & \tiny 87 & \tiny 87 & \tiny 82 & \tiny 75 & \tiny 73 & \tiny 64 & \tiny 38 & \tiny 47 & \tiny 77 \\
\tiny Per-emb-5 & \tiny 03:31:20.942 & \tiny +30:45:30.263 & \tiny IRAS~03282+3035 & \tiny Y & \tiny 32 & \tiny 1.2 & \tiny 126 & \tiny -56 & \tiny 125 & \tiny (1) & \tiny 39 & \tiny 86 & \tiny 75 & \tiny 80 & \tiny 73 & \tiny 80 & \tiny 84 & \tiny 80 & \tiny 73 & \tiny 81 \\
\tiny Per-emb-6 & \tiny 03:33:14.404 & \tiny +31:07:10.715 & \tiny ... & \tiny N & \tiny 52 & \tiny 2.9 & \tiny 50 & \tiny -109 & \tiny 60 & \tiny (1) & \tiny 48 & \tiny 12 & \tiny 14 & \tiny 11 & \tiny 11 & \tiny 11 & \tiny 9 & \tiny 11 & \tiny 15 & \tiny 16 \\
\tiny Per-emb-8 & \tiny 03:44:43.982 & \tiny +32:01:35.210 & \tiny ... & \tiny Y & \tiny 43 & \tiny 0.7 & \tiny 15 & \tiny -165 & \tiny 15 & \tiny (3) & \tiny 65 & \tiny 50 & \tiny 47 & \tiny 47 & \tiny 48 & \tiny 45 & \tiny 50 & \tiny 50 & \tiny 88 & \tiny 62 \\
\tiny Per-emb-9 & \tiny 03:29:51.832 & \tiny +31:39:05.905 & \tiny IRAS~03267+3128,~Perseus5 & \tiny N & \tiny 36 & \tiny 0.8 & \tiny 63 & \tiny -125 & \tiny 59 & \tiny (1) & \tiny 54 & \tiny 5 & \tiny 2 & \tiny 1 & \tiny 1 & \tiny 3 & \tiny 77 & \tiny 14 & \tiny 27 & \tiny 39 \\
\tiny Per-emb-10 & \tiny 03:33:16.424 & \tiny +31:06:52.063 & \tiny ... & \tiny N & \tiny 30 & \tiny 3.8 & \tiny -134 & \tiny 57 & \tiny 52 & \tiny (1) & \tiny 48 & \tiny 4 & \tiny 5 & \tiny 2 & \tiny 2 & \tiny 2 & \tiny 1 & \tiny 3 & \tiny 24 & \tiny 8 \\
\tiny Per-emb-11,O1 & \tiny 03:43:57.065 & \tiny +32:03:04.788 & \tiny IC348MMS & \tiny Y & \tiny 30 & \tiny 1.6 & \tiny -17 & \tiny 161 & \tiny 162 & \tiny (1) & \tiny 134 & \tiny 27 & \tiny 27 & \tiny 38 & \tiny 40 & \tiny 38 & \tiny 37 & \tiny 61 & \tiny 78 & \tiny 85 \\
\tiny Per-emb-11,O2 & \tiny 03:43:57.688 & \tiny +32:03:09.975 & \tiny IC348MMS & \tiny Y & \tiny 30 & \tiny 1.9 & \tiny 36 & \tiny ... & \tiny 36 & \tiny (3),(7) & \tiny 134 & \tiny 82 & \tiny 81 & \tiny 88 & \tiny 86 & \tiny 88 & \tiny 89 & \tiny 7 & \tiny 48 & \tiny 41 \\
\tiny Per-emb-12 & \tiny 03:29:10.537 & \tiny +31:13:30.925 & \tiny NGC~1333~IRAS4A & \tiny Y & \tiny 29 & \tiny 4.6 & \tiny -145 & \tiny 35 & \tiny 35 & \tiny (2) & \tiny 128 & \tiny 87 & \tiny 87 & \tiny 85 & \tiny 85 & \tiny 0 & \tiny 9 & \tiny 80 & \tiny 82 & \tiny 15 \\
\tiny Per-emb-13,O1 & \tiny 03:29:12.016 & \tiny +31:13:08.031 & \tiny NGC~1333~IRAS4B & \tiny Y & \tiny 28 & \tiny 7.1 & \tiny 180 & \tiny 0 & \tiny 180 & \tiny (2) & \tiny 130 & \tiny 50 & \tiny 50 & \tiny 48 & \tiny 46 & \tiny 35 & \tiny 26 & \tiny 45 & \tiny 47 & \tiny 21 \\
\tiny Per-emb-13,O2 & \tiny 03:29:12.842 & \tiny +31:13:06.893 & \tiny NGC~1333~IRAS4B$^\prime$ & \tiny Y & \tiny 28 & \tiny 7.9 & \tiny -90 & \tiny 90 & \tiny 90 & \tiny (3) & \tiny 130 & \tiny 40 & \tiny 40 & \tiny 41 & \tiny 44 & \tiny 49 & \tiny 64 & \tiny 45 & \tiny 43 & \tiny 70 \\
\tiny Per-emb-15 & \tiny 03:29:04.055 & \tiny +31:14:46.237 & \tiny RNO15-FIR & \tiny N & \tiny 36 & \tiny 3.1 & \tiny 145 & \tiny -35 & \tiny 145 & \tiny (2) & \tiny 42 & \tiny 77 & \tiny 77 & \tiny 82 & \tiny 87 & \tiny 4 & \tiny 61 & \tiny 14 & \tiny 12 & \tiny 56 \\
\tiny Per-emb-16 & \tiny 03:43:50.978 & \tiny +32:03:24.101 & \tiny ... & \tiny Y & \tiny 39 & \tiny 1.6 & \tiny 14 & \tiny -173 & \tiny 11 & \tiny (1) & \tiny 77 & \tiny 67 & \tiny 82 & \tiny 85 & \tiny 84 & \tiny 79 & \tiny 75 & \tiny 78 & \tiny 73 & \tiny 66 \\
\tiny Per-emb-17 & \tiny 03:27:39.104 & \tiny +30:13:03.078 & \tiny ... & \tiny Y & \tiny 59 & \tiny 0.5 & \tiny -127 & \tiny 60 & \tiny 57 & \tiny (1) & \tiny 146 & \tiny 89 & \tiny 90 & \tiny 81 & \tiny 82 & \tiny 86 & \tiny 83 & \tiny 83 & \tiny 65 & \tiny 90 \\
\tiny Per-emb-18 & \tiny 03:29:11.258 & \tiny +31:18:31.073 & \tiny NGC~1333~IRAS7 & \tiny Y & \tiny 59 & \tiny 1.3 & \tiny -30 & \tiny 150 & \tiny 150 & \tiny (3) & \tiny 20 & \tiny 50 & \tiny 73 & \tiny 3 & \tiny 7 & \tiny 75 & \tiny 74 & \tiny 66 & \tiny 63 & \tiny 51 \\
\tiny Per-emb-19 & \tiny 03:29:23.498 & \tiny +31:33:29.173 & \tiny ... & \tiny N & \tiny 60 & \tiny 1.0 & \tiny -32 & \tiny 148 & \tiny 148 & \tiny (1),(8) & \tiny 27 & \tiny 59 & \tiny 57 & \tiny 77 & \tiny 17 & \tiny 17 & \tiny 24 & \tiny 42 & \tiny 55 & \tiny 53 \\
\tiny Per-emb-20 & \tiny 03:27:43.276 & \tiny +30:12:28.781 & \tiny L1455-IRS4 & \tiny N & \tiny 65 & \tiny 1.6 & \tiny -61 & \tiny 112 & \tiny 115 & \tiny (1) & \tiny 58 & \tiny 58 & \tiny 18 & \tiny 15 & \tiny 26 & \tiny 27 & \tiny 29 & \tiny 28 & \tiny 6 & \tiny 32 \\
\tiny Per-emb-21 & \tiny 03:29:10.668 & \tiny +31:18:20.191 & \tiny ... & \tiny Y & \tiny 45 & \tiny 1.6 & \tiny 48 & \tiny -132 & \tiny 48 & \tiny (3) & \tiny 20 & \tiny 28 & \tiny 15 & \tiny 35 & \tiny 25 & \tiny 3 & \tiny 4 & \tiny 22 & \tiny 15 & \tiny 28 \\
\tiny Per-emb-22 & \tiny 03:25:22.410 & \tiny +30:45:13.254 & \tiny L1448-IRS2 & \tiny Y & \tiny 43 & \tiny 1.1 & \tiny -62 & \tiny 118 & \tiny 118 & \tiny (1),(8) & \tiny 61 & \tiny 57 & \tiny 43 & \tiny 37 & \tiny 37 & \tiny 37 & \tiny 37 & \tiny 36 & \tiny 26 & \tiny 29 \\
\tiny Per-emb-23 & \tiny 03:29:17.211 & \tiny +31:27:46.302 & \tiny ASR~30 & \tiny N & \tiny 42 & \tiny 1.0 & \tiny -125 & \tiny 61 & \tiny 58 & \tiny (1) & \tiny 138 & \tiny 79 & \tiny 17 & \tiny 17 & \tiny 45 & \tiny 43 & \tiny 46 & \tiny 85 & \tiny 56 & \tiny 38 \\
\tiny Per-emb-24 & \tiny 03:28:45.297 & \tiny +31:05:41.693 & \tiny ... & \tiny N & \tiny 67 & \tiny 0.9 & \tiny -103 & \tiny 93 & \tiny 85 & \tiny (1) & \tiny 54 & \tiny 31 & \tiny 29 & \tiny 27 & \tiny 26 & \tiny 23 & \tiny 56 & \tiny 59 & \tiny 33 & \tiny 64 \\
\tiny Per-emb-25 & \tiny 03:26:37.511 & \tiny +30:15:27.813 & \tiny ... & \tiny N & \tiny 61 & \tiny 0.4 & \tiny -78 & \tiny 107 & \tiny 104 & \tiny (1) & \tiny 61 & \tiny 43 & \tiny 31 & \tiny 45 & \tiny 64 & \tiny 47 & \tiny 48 & \tiny 10 & \tiny 84 & \tiny 43 \\
\tiny Per-emb-26 & \tiny 03:25:38.875 & \tiny +30:44:05.283 & \tiny L1448C,~L1448-mm & \tiny Y & \tiny 47 & \tiny 1.8 & \tiny -21 & \tiny 165 & \tiny 162 & \tiny (1) & \tiny 130 & \tiny 32 & \tiny 39 & \tiny 36 & \tiny 35 & \tiny 39 & \tiny 33 & \tiny 39 & \tiny 70 & \tiny 73 \\
\tiny Per-emb-27,O1 & \tiny 03:28:55.569 & \tiny +31:14:37.022 & \tiny NGC~1333~IRAS2A & \tiny Y & \tiny 69 & \tiny 1.7 & \tiny -156 & \tiny 4 & \tiny 14 & \tiny (2) & \tiny 125 & \tiny 69 & \tiny 80 & \tiny 84 & \tiny 79 & \tiny 46 & \tiny 22 & \tiny 27 & \tiny 57 & \tiny 7 \\
\tiny Per-emb-27,O2 & \tiny 03:28:55.563 & \tiny +31:14:36.408 & \tiny NGC~1333~IRAS2A & \tiny Y & \tiny 69 & \tiny 1.7 & \tiny -77 & \tiny 105 & \tiny 104 & \tiny (2) & \tiny 125 & \tiny 21 & \tiny 10 & \tiny 6 & \tiny 11 & \tiny 44 & \tiny 68 & \tiny 63 & \tiny 33 & \tiny 84 \\
\tiny Per-emb-28 & \tiny 03:43:51.008 & \tiny +32:03:08.042 & \tiny ... & \tiny Y & \tiny 45 & \tiny 1.8 & \tiny 112 & \tiny -68 & \tiny 112 & \tiny (3) & \tiny 77 & \tiny 35 & \tiny 3 & \tiny 6 & \tiny 5 & \tiny 0 & \tiny 4 & \tiny 21 & \tiny 28 & \tiny 35 \\
\tiny Per-emb-29 & \tiny 03:33:17.877 & \tiny +31:09:31.817 & \tiny B1-c & \tiny N & \tiny 48 & \tiny 2.7 & \tiny 133 & \tiny -50 & \tiny 132 & \tiny (1) & \tiny 7 & \tiny 55 & \tiny 58 & \tiny 57 & \tiny 62 & \tiny 48 & \tiny 80 & \tiny 7 & \tiny 20 & \tiny 88 \\
\tiny Per-emb-33,O1 & \tiny 03:25:36.380 & \tiny +30:45:14.723 & \tiny L1448IRS3B,~L1448N & \tiny Y & \tiny 57 & \tiny 4.7 & \tiny -58 & \tiny 122 & \tiny 122 & \tiny (3),(4) & \tiny 127 & \tiny 5 & \tiny 12 & \tiny 9 & \tiny 12 & \tiny 13 & \tiny 9 & \tiny 10 & \tiny 30 & \tiny 33 \\
\tiny Per-emb-33,O2 & \tiny 03:25:36.499 & \tiny +30:45:21.880 & \tiny L1448IRS3B,~L1448N & \tiny Y & \tiny 57 & \tiny 4.8 & \tiny 38 & \tiny -142 & \tiny 38 & \tiny (3),(4) & \tiny 127 & \tiny 89 & \tiny 72 & \tiny 75 & \tiny 72 & \tiny 71 & \tiny 69 & \tiny 74 & \tiny 54 & \tiny 51 \\
\tiny Per-emb-33,O3 & \tiny 03:25:35.669 & \tiny +30:45:34.110 & \tiny L1448IRS3B,~L1448N & \tiny Y & \tiny 57 & \tiny 4.3 & \tiny -52 & \tiny 128 & \tiny 128 & \tiny (3),(4) & \tiny 130 & \tiny 2 & \tiny 19 & \tiny 19 & \tiny 23 & \tiny 24 & \tiny 21 & \tiny 19 & \tiny 36 & \tiny 39 \\
\tiny Per-emb-35,O1 & \tiny 03:28:37.091 & \tiny +31:13:30.788 & \tiny NGC~1333~IRAS1 & \tiny Y & \tiny 103 & \tiny 0.6 & \tiny -57 & \tiny 123 & \tiny 123 & \tiny (1),(8) & \tiny 32 & \tiny 89 & \tiny 68 & \tiny 73 & \tiny 86 & \tiny 86 & \tiny 89 & \tiny 85 & \tiny 67 & \tiny 78 \\
\tiny Per-emb-35,O2 & \tiny 03:28:37.219 & \tiny +31:13:31.751 & \tiny NGC~1333~IRAS1 & \tiny Y & \tiny 103 & \tiny 0.6 & \tiny 169 & \tiny -11 & \tiny 169 & \tiny (1),(8) & \tiny 32 & \tiny 43 & \tiny 22 & \tiny 27 & \tiny 48 & \tiny 49 & \tiny 46 & \tiny 39 & \tiny 67 & \tiny 32 \\
\tiny Per-emb-36 & \tiny 03:28:57.374 & \tiny +31:14:15.765 & \tiny NGC~1333~IRAS2B & \tiny Y & \tiny 106 & \tiny 1.6 & \tiny -156 & \tiny ... & \tiny 24 & \tiny (2),(7) & \tiny 134 & \tiny 70 & \tiny 85 & \tiny 85 & \tiny 89 & \tiny 7 & \tiny 9 & \tiny 76 & \tiny 67 & \tiny 4 \\
\tiny Per-emb-37 & \tiny 03:29:18.965 & \tiny +31:23:14.304 & \tiny ... & \tiny Y & \tiny 22 & \tiny 0.8 & \tiny -139 & \tiny 34 & \tiny 38 & \tiny (1) & \tiny 30 & \tiny 7 & \tiny 7 & \tiny 22 & \tiny 12 & \tiny 4 & \tiny 8 & \tiny 8 & \tiny 20 & \tiny 17 \\
\tiny Per-emb-40 & \tiny 03:33:16.669 & \tiny +31:07:54.902 & \tiny B1-a & \tiny Y & \tiny 132 & \tiny 2.0 & \tiny 101 & \tiny -79 & \tiny 101 & \tiny (1),(9) & \tiny 44 & \tiny 57 & \tiny 33 & \tiny 54 & \tiny 54 & \tiny 51 & \tiny 51 & \tiny 34 & \tiny 26 & \tiny 57 \\
\tiny Per-emb-41 & \tiny 03:33:20.341 & \tiny +31:07:21.355 & \tiny B1-b & \tiny Y & \tiny 157 & \tiny 4.1 & \tiny -150 & \tiny 30 & \tiny 30 & \tiny (3) & \tiny 125 & \tiny 85 & \tiny 85 & \tiny 84 & \tiny 83 & \tiny 84 & \tiny 80 & \tiny 82 & \tiny 45 & \tiny 14 \\
\tiny Per-emb-42 & \tiny 03:25:39.135 & \tiny +30:43:57.909 & \tiny L1448C-S & \tiny Y & \tiny 163 & \tiny 1.9 & \tiny 43 & \tiny -137 & \tiny 43 & \tiny (3) & \tiny 130 & \tiny 87 & \tiny 80 & \tiny 83 & \tiny 84 & \tiny 80 & \tiny 85 & \tiny 80 & \tiny 49 & \tiny 46 \\
\tiny Per-emb-44 & \tiny 03:29:03.766 & \tiny +31:16:03.810 & \tiny SVS13A & \tiny Y & \tiny 188 & \tiny 3.0 & \tiny 120 & \tiny -40 & \tiny 130 & \tiny (2) & \tiny 13 & \tiny 63 & \tiny 72 & \tiny 70 & \tiny 65 & \tiny 21 & \tiny 76 & \tiny 3 & \tiny 3 & \tiny 71 \\
\tiny Per-emb-46 & \tiny 03:28:00.415 & \tiny +30:08:01.013 & \tiny ... & \tiny N & \tiny 221 & \tiny 0.8 & \tiny -49 & \tiny 131 & \tiny 131 & \tiny (10) & \tiny 10 & \tiny 59 & \tiny 62 & \tiny 64 & \tiny 21 & \tiny 24 & \tiny 20 & \tiny 15 & \tiny 20 & \tiny 16 \\
\tiny Per-emb-49 & \tiny 03:29:12.953 & \tiny +31:18:14.289 & \tiny ... & \tiny Y & \tiny 239 & \tiny 2.3 & \tiny -153 & \tiny 27 & \tiny 27 & \tiny (3) & \tiny 20 & \tiny 7 & \tiny 16 & \tiny 15 & \tiny 4 & \tiny 18 & \tiny 17 & \tiny 9 & \tiny 6 & \tiny 7 \\
\tiny Per-emb-50 & \tiny 03:29:07.768 & \tiny +31:21:57.128 & \tiny ... & \tiny N & \tiny 128 & \tiny 0.7 & \tiny -83 & \tiny 112 & \tiny 104 & \tiny (1) & \tiny 120 & \tiny 16 & \tiny 16 & \tiny 19 & \tiny 16 & \tiny 62 & \tiny 66 & \tiny 67 & \tiny 48 & \tiny 84 \\
\tiny Per-emb-53 & \tiny 03:47:41.591 & \tiny +32:51:43.672 & \tiny B5-IRS1 & \tiny N & \tiny 287 & \tiny 0.8 & \tiny 52 & \tiny -114 & \tiny 59 & \tiny (1) & \tiny 26 & \tiny 33 & \tiny 30 & \tiny 13 & \tiny 43 & \tiny 34 & \tiny 46 & \tiny 42 & \tiny 74 & \tiny 18 \\
\tiny Per-emb-55 & \tiny 03:44:43.298 & \tiny +32:01:31.223 & \tiny IRAS~03415+3152 & \tiny Y & \tiny 309 & \tiny 0.5 & \tiny 115 & \tiny -65 & \tiny 115 & \tiny (3) & \tiny 65 & \tiny 50 & \tiny 53 & \tiny 53 & \tiny 52 & \tiny 55 & \tiny 56 & \tiny 50 & \tiny 8 & \tiny 38 \\
\tiny Per-emb-56 & \tiny 03:47:05.450 & \tiny +32:43:08.240 & \tiny IRAS~03439+3233 & \tiny N & \tiny 312 & \tiny 0.4 & \tiny 145 & \tiny -35 & \tiny 145 & \tiny (10) & \tiny 54 & \tiny 89 & \tiny 81 & \tiny 78 & \tiny 15 & \tiny 82 & \tiny 86 & \tiny 81 & \tiny 66 & \tiny 76 \\
\tiny Per-emb-57 & \tiny 03:29:03.331 & \tiny +31:23:14.573 & \tiny ... & \tiny N & \tiny 313 & \tiny 0.4 & \tiny 145 & \tiny 147 & \tiny 146 & \tiny (1),(11) & \tiny 135 & \tiny 11 & \tiny 7 & \tiny 11 & \tiny 34 & \tiny 75 & \tiny 73 & \tiny 68 & \tiny 6 & \tiny 54 \\
\tiny Per-emb-58 & \tiny 03:28:58.422 & \tiny +31:22:17.481 & \tiny ... & \tiny N & \tiny 322 & \tiny 1.2 & \tiny -13 & \tiny ... & \tiny 167 & \tiny (1),(7) & \tiny 135 & \tiny 32 & \tiny 29 & \tiny 32 & \tiny 54 & \tiny 58 & \tiny 55 & \tiny 54 & \tiny 14 & \tiny 34 \\
\tiny Per-emb-61 & \tiny 03:44:21.357 & \tiny +31:59:32.514 & \tiny ... & \tiny N & \tiny 371 & \tiny 0.7 & \tiny 15 & \tiny -165 & \tiny 15 & \tiny (1),(9) & \tiny 134 & \tiny 61 & \tiny 61 & \tiny 66 & \tiny 75 & \tiny 83 & \tiny 86 & \tiny 84 & \tiny 41 & \tiny 62 \\
\tiny Per-emb-62 & \tiny 03:44:12.977 & \tiny +32:01:35.419 & \tiny ... & \tiny N & \tiny 378 & \tiny 0.4 & \tiny -155 & \tiny 24 & \tiny 24 & \tiny (1) & \tiny 132 & \tiny 72 & \tiny 76 & \tiny 63 & \tiny 79 & \tiny 75 & \tiny 78 & \tiny 84 & \tiny 76 & \tiny 52 \\
\tiny SVS13B & \tiny 03:29:03.078 & \tiny +31:15:51.740 & \tiny ... & \tiny Y & \tiny 20 & \tiny 2.7 & \tiny ... & \tiny -20 & \tiny 170 & \tiny (3),(7) & \tiny 14 & \tiny 24 & \tiny 32 & \tiny 30 & \tiny 46 & \tiny 22 & \tiny 36 & \tiny 41 & \tiny 37 & \tiny 31 \\
\tiny SVS13C & \tiny 03:29:01.970 & \tiny +31:15:38.053 & \tiny ... & \tiny Y & \tiny 21 & \tiny 2.5 & \tiny -172 & \tiny 8 & \tiny 8 & \tiny (2) & \tiny 14 & \tiny 6 & \tiny 14 & \tiny 30 & \tiny 28 & \tiny 39 & \tiny 18 & \tiny 61 & \tiny 55 & \tiny 13 \\
\tiny B1-bN & \tiny 03:33:21.209 & \tiny +31:07:43.665 & \tiny ... & \tiny Y & \tiny 14.7 & \tiny 4.9 & \tiny 90 & \tiny ... & \tiny 90 & \tiny (3),(7) & \tiny 128 & \tiny 38 & \tiny 37 & \tiny 38 & \tiny 39 & \tiny 38 & \tiny 43 & \tiny 38 & \tiny 45 & \tiny 76 \\
\tiny B1-bS & \tiny 03:33:21.355 & \tiny +31:07:26.372 & \tiny ... & \tiny Y & \tiny 17.7 & \tiny 5.8 & \tiny 112 & \tiny -68 & \tiny 120 & \tiny (3) & \tiny 125 & \tiny 5 & \tiny 5 & \tiny 6 & \tiny 7 & \tiny 6 & \tiny 10 & \tiny 7 & \tiny 15 & \tiny 46 \\
\enddata
\end{deluxetable*}

\begin{deluxetable*}{lccl@{}ccccccccccccccccc}
\tabletypesize{\tiny}
\tablenum{2}
\tablecaption{Source Information and Measured Position Angles \label{tab:angles}}
\tablehead{\colhead{\tiny Source} & \colhead{\tiny RA\tablenotemark{b}} & \colhead{\tiny DEC\tablenotemark{b}} & \colhead{\tiny Other Names\tablenotemark{c}} & \colhead{\tiny Multiple} & \colhead{\tiny $T_{\rm{bol}}$} & \colhead{\tiny \od} & \colhead{\tiny Blue PA} & \colhead{\tiny Red PA} & \colhead{\tiny \pao} & \colhead{\tiny Ref/Info} & \colhead{\tiny \pafilf} & \colhead{\tiny \gaf}  & \colhead{\tiny \pafone} & \colhead{\tiny \paftwo} & \colhead{\tiny \pafthree} & \colhead{\tiny \paffour} & \colhead{\tiny \paffive} & \colhead{\tiny \pafsix} & \colhead{\tiny $\gamma_{\mbox{se,S}}$} & \colhead{\tiny $\gamma_{\mbox{se,L}}$}\\
\colhead{\tiny Name\tablenotemark{a}} & \colhead{\tiny (J2000)} & \colhead{\tiny (J2000)} & & \colhead{\tiny (Y/N)} & \colhead{\tiny (K)} & \tiny ($\times 10^3$) & \colhead{\tiny ($^\circ$)} & \colhead{\tiny ($^\circ$)} & \colhead{\tiny ($^\circ$)} & & 
\colhead{\tiny ($^\circ$)} & \colhead{\tiny ($^\circ$)} & \colhead{\tiny ($^\circ$)} & \colhead{\tiny ($^\circ$)} & \colhead{\tiny ($^\circ$)} & \colhead{\tiny ($^\circ$)} & \colhead{\tiny ($^\circ$)} & \colhead{\tiny ($^\circ$)} & \colhead{\tiny ($^\circ$)} & \colhead{\tiny ($^\circ$)}
}
\startdata
\tiny L1448IRS2E & \tiny 03:25:25.660 & \tiny +30:44:56.695 & \tiny ... & \tiny N & \tiny 15 & \tiny 2.6 & \tiny ... & \tiny 165 & \tiny 165 & \tiny (5),(7) & \tiny 62 & \tiny 77 & \tiny 87 & \tiny 87 & \tiny 86 & \tiny 83 & \tiny 84 & \tiny 81 & \tiny 73 & \tiny 76 \\
\tiny L1451-MMS & \tiny 03:25:10.245 & \tiny +30:23:55.059 & \tiny ... & \tiny N & \tiny 15 & \tiny 0.9 & \tiny 11 & \tiny -169 & \tiny 11 & \tiny (6) & \tiny 123 & \tiny 68 & \tiny 71 & \tiny 37 & \tiny 71 & \tiny 60 & \tiny 61 & \tiny 34 & \tiny 54 & \tiny 53 \\
\tiny Per-bolo-58 & \tiny 03:29:25.464 & \tiny +31:28:14.880 & \tiny ... & \tiny N\tablenotemark{d} & \tiny 15 & \tiny 0.9 & \tiny 87 & \tiny -93 & \tiny 87 & \tiny (1) & \tiny 56 & \tiny 32 & \tiny 40 & \tiny 41 & \tiny 65 & \tiny 67 & \tiny 72 & \tiny 52 & \tiny 41 & \tiny 67
\enddata
\tablecomments{(1) Our study; measured by connecting outflows to continuum peaks, (2) \citet{Plunkett2013}, (3) \citet{Lee2016}, (4) \citet{Lee2015}, (5) \citet{Chen2010}, measured manually by our study, (6) \citet{Pineda2011}, measured manually by our study, (7) only one outflow lobe detected in the cited study, (8) outflow PA fit only using blue lobe, (9) outflow PA fit only using red lobe, (10) our study; PA measured manually; (11) red and blue lobe are both in same quadrant; we consider this as a single outflow that may be in the plane of the sky.}
\tablenotetext{a}{Names including O1, O2, and O3 are sources with multiple outflows.}
\tablenotetext{b}{RA and DEC positions are from \citet{Tobin2016}. In the case where a close binary is unresolved by the SMA, we pick the brightest \citet{Tobin2016} protostar for the source of the emission.}
\tablenotetext{c}{Alternate names are taken from \citet{Tobin2016}.}
\tablenotetext{d}{This sources was not detected in \citet{Tobin2016}.}
\end{deluxetable*}

\global\pdfpageattr\expandafter{\the\pdfpageattr/Rotate 90}
\end{turnpage}
\end{document}